\documentclass[12pt]{article}
\usepackage{graphicx,amssymb,amsmath}
\usepackage{epsfig}
\usepackage{color,graphicx}
\usepackage{cite}
\usepackage{amssymb,amsmath}
\newcommand{\rc}{\nonumber\\}
\newcommand{\be}{\begin{equation}}
\newcommand{\ee}{\end{equation}}
\newcommand{\bea}{\begin{eqnarray}}
\newcommand{\eea}{\end{eqnarray}}
\newcommand{\beas}{\begin{eqnarray*}}
\newcommand{\eeas}{\end{eqnarray*}}
\newcommand{\beq}{\begin{equation}}
\newcommand{\eeq}{\end{equation}}
\newcommand{\bear}{\begin{eqnarray}}
\newcommand{\eear}{\end{eqnarray}}
\def\ie{\hbox{\emph{i.e.}}}
\numberwithin{equation}{section}

\begin{document}

\begin{flushright}
HIP-2014-26/TH
\end{flushright}

\begin{center}

\centerline{\Large {\bf Flux and Hall states in ABJM with dynamical flavors}}

\vspace{8mm}

\renewcommand\thefootnote{\mbox{$\fnsymbol{footnote}$}}
Yago Bea,${}^{1}$\footnote{yago.bea@fpaxp1.usc.es}
Niko Jokela,${}^{2,3}$\footnote{niko.jokela@helsinki.fi}
Matthew Lippert,${}^{4}$\footnote{M.S.Lippert@uva.nl} \\
Alfonso V. Ramallo,${}^{1}$\footnote{alfonso@fpaxp1.usc.es} and
Dimitrios Zoakos${}^5$\footnote{dimitrios.zoakos@fc.up.pt}

\vspace{6mm}
${}^1${\small \sl Departamento de F\'isica de Part\'iculas and  Instituto Galego de F\'isica de Altas Enerx\'ias}\\ 
{\small \sl  Universidade de Santiago de Compostela} \\ 
{\small \sl E-15782 Santiago de Compostela, Spain}

\vspace{3mm}
${}^2${\small \sl Department of Physics and ${}^3$Helsinki Institute of Physics} \\
{\small \sl P.O. Box 64, FIN-00014 University of Helsinki, Finland}

\vspace{3mm}
${}^4${\small \sl Institute for Theoretical Physics} \\
{\small \sl University of Amsterdam} \\
{\small \sl  1090GL Amsterdam, Netherlands }

\vspace{3mm}
${}^5${\small \sl Centro de F\'isica do Porto 
 and Departamento de F\'isica e Astronomia} \\
{\small \sl Faculdade de Ci\^encias da Universidade do Porto}\\
{\small \sl Rua do Campo Alegre 687, 4169-007 Porto, Portugal} 

\end{center}

\vspace{8mm}

\setcounter{footnote}{0}
\renewcommand\thefootnote{\mbox{\arabic{footnote}}}

\begin{abstract}
\noindent
We study the physics of probe D6-branes with quantized internal worldvolume flux in the ABJM background with unquenched massless flavors. This flux breaks parity in the (2+1)-dimensional gauge theory and allows quantum Hall states. Parity breaking is also explicitly demonstrated via the helicity dependence of the meson spectrum.
We obtain general expressions for the conductivities, both in the gapped Minkowski embeddings and in the compressible black hole ones. These conductivities depend on the flux
and contain a contribution from the dynamical flavors which can be regarded as an effect of intrinsic disorder due to quantum fluctuations of the fundamentals. We present an explicit, analytic family of supersymmetric solutions with nonzero charge density, electric, and magnetic fields.

\end{abstract}

\newpage
\tableofcontents

\section{Introduction and motivation}\label{intro}

The quantum Hall effect (QHE) is a fascinating phenomenon in gapped (2+1)-dimensional systems with broken parity symmetry. When electrons are confined in a heterojunction at low temperature and strong magnetic fields, the response to an applied electric field displays a striking behavior: the conductivity in the direction of the electric field vanishes, while the transverse conductivity is quantized and given by $(e^2/h)\nu$, where $\nu$ is the filling fraction, defined as the ratio of the charge density to the magnetic flux. In the integer quantum Hall effect (IQHE) $\nu\in {\mathbb Z}$, whereas $\nu$ is a rational number  in the fractional quantum Hall effect (FQHE).

Since its discovery more than thirty years ago, the QHE has been the subject of intense research.  Nevertheless, some aspects of the FQHE involve strongly-coupled  dynamics and are still not fully understood. The holographic AdS/CFT duality has proven to be a powerful tool in the study of quantum matter in the strongly-coupled regime, since it provides answers to difficult field theory questions by using classical gravitational theories in higher dimensions.  Therefore, it is quite natural to explore the possibility of constructing holographic models of the (F)QHE and to extract properties that are very difficult to obtain via weakly-coupled many-body field theory.

In recent years, two types of holographic models of the QHE have been proposed. The  first class consists of bottom-up models in Einstein-Maxwell-axio-dilaton theories \cite{KeskiVakkuri:2008eb, Goldstein:2009cv,Goldstein:2010aw, Bayntun:2010nx,Gubankova:2010rc}. These models are endowed with an $SL(2,{\mathbb Z})$ duality and, as a consequence, they capture some observed features of QH physics. However, it is very difficult to engineer these types of models to have a mass gap; \cite{Lippert:2014jma} is so far the only example of a gapped model in this class. 

The second approach to holographically realize the QHE makes use of top-down D-brane constructions \cite{Bergman:2010gm,Jokela:2011eb,Kristjansen:2012ny}, in which a (2+1)-dimensional gauge theory  with fermions in the fundamental representation is modeled  by a suitable D$p$-D$q$ brane intersection. The limit in which the D$q$-brane is treated as a probe in the D$p$-brane background corresponds in the field theory dual to the so-called quenched approximation in which loops of fundamental fermions are neglected. In this approach, the worldvolume theory of the probe brane encodes the physics of the fermions.  Generically, the probe brane crosses the horizon, yielding a black hole embedding, which is dual to a gapless metallic state.  The quantum Hall state is realized holographically as a Minkowski embedding, in which the brane ends smoothly above the black hole horizon. The distance from the horizon at which the probe caps off determines the mass gap.

In this paper we will follow the top-down probe-brane approach and construct quantum Hall states in the ABJM theory with unquenched massless flavors. The unflavored ABJM model is a 
$U(N)\times U(N)$ Chern-Simons gauge theory in 2+1 dimensions with levels $(k,-k)$ and bifundamental matter fields \cite{Aharony:2008ug}.  In string theory, the ABJM theory is realized as the low-energy limit of multiple M2-branes at a ${\mathbb C}^4/{\mathbb Z}_k$ singularity.  When $N$ and $k$ are large, this theory admits a supergravity description, preserving 24 supersymmetries,  in terms of a 
$AdS_4\times {\mathbb C}{\mathbb P}^3$ geometry with fluxes in type IIA ten-dimensional supergravity.    Due to its high degree of supersymmetry, the ABJM theory is one of the models where the AdS/CFT correspondence has been tested with great precision. Since the boundary theory is conformally invariant and the bulk metric therefore has an $AdS$ factor, the gauge/gravity dictionary is firmly established.

The ABJM model can be generalized by adding flavors, \ie, fields transforming in the fundamental representations $(N,1)$ and $(1,N)$ of the $U(N)\times U(N)$ gauge group, which we will refer to as ``quarks" in analogy with the terminology of holographic QCD. In the holographic setup, these flavors are due to D6-branes extended in $AdS_4$ and wrapping an ${\mathbb R}{\mathbb P}^3$ cycle inside
${\mathbb C}{\mathbb P}^3$ \cite{Hohenegger:2009as,Gaiotto:2009tk}.  This configuration preserves ${\cal N}=3$ supersymmetry.  The quenched approximation of these holographic quarks, where the D6-branes are treated as probes, has been studied in \cite{Hikida:2009tp,Jensen:2010vx,Ammon:2009wc, Alanen:2009cn, Zafrir:2012yg}.

However, it is possible to go beyond the quenched approximation and include the backreaction of the D6-branes; there are simple analytic geometries which encode the dynamics of the flavors in the Veneziano limit\cite{Veneziano:1976wm}.  Here we will employ the solution found in \cite{Conde:2011sw,Jokela:2012dw} by  smearing the D6-branes.\footnote{See \cite{Nunez:2010sf} and references therein for a review.} This smearing technique is applicable when the number $N_f$ of flavor branes is large and can be continuously distributed in the internal space, which changes the flavor group from $U(N_f)$ to $U(1)^{N_f}$. For massless flavors the result is simply a metric which differs from the unflavored one merely by constant squashing factors.  The construction was generalized in \cite{Bea:2013jxa} to the backreaction of massive flavors.  These squashing factors depend on $N_f$ and encode the effects of dynamical flavor loops.

In this paper, we want to engineer quantum Hall states in the flavored ABJM theory.  Such Hall states are only possible if parity is broken, which can be accomplished by turning on an appropriate internal flux on the D6-brane worldvolume. However, treating the backreaction of this internal flux is quite challenging.  For now, we will start with a single quenched massive quark in the background of $N_f$ unquenched massless quarks, a system analyzed in \cite{Jokela:2013qya, Jokela:2012dw}.  We then will turn on a parity-breaking internal flux on the worldvolume of this probe D6-brane.

In the presence of this internal flux, the Wess-Zumino term of the probe action contains the term $\int \hat C_1\wedge F^3$, where $\hat C_1$ is the pullback of the RR potential one-form. In the ABJM background $C_1$ has only internal components. Therefore, after integrating over the internal directions, we are left with an axionic term $F\wedge F$ along $AdS_4$, which indeed breaks parity and corresponds to a Chern-Simons term on the boundary. 

Even in the probe limit, choosing a consistent ansatz for this internal flux, which must also be quantized appropriately, is not obvious.   We can, however, take a cue from the ABJ model \cite{Aharony:2008gk}, \ie,  the $U(N+M)_{k}\times U(N)_{-k}$ Chern-Simons matter theory, which can be engineered in string theory by adding fractional D2-branes to the ABJM setup.  The corresponding gravity dual can be obtained from the ABJM solution by turning on a flat Neveu-Schwarz $B_2$ field proportional to the K\"ahler form of ${\mathbb C}{\mathbb P}^3$.  The pullback of this parity-breaking $B_2$ on a probe D6-brane can alternately be viewed as a worldvolume gauge field flux.  Inspired by this example, we will generalize this ABJ solution into an ansatz for the case with no background $B_2$ field and only a probe worldvolume flux, but with backreacted massless flavors.

Equipped with this ansatz for the internal gauge flux, we will show that, indeed, there are quantum Hall states in this setup.  From the QH perspective, one can regard the effects of the massless, backreacted quarks as representing intrinsic disorder due to the quantum fluctuations of the massive quark. We will compute the contribution of these fluctuations to the conductivities in the form of an integral extended in the holographic direction, from the tip of the brane to the $AdS$ boundary. 

Surprisingly, we will find a very special family of explicit, supersymmetric, gapped QH solutions at zero temperature.  These BPS solutions have nonzero charge density and equal electric and magnetic fields, and we can compute the Hall conductivity, including the effects of quark loops, analytically.

The rest of this paper is organized as follows. In Section \ref{Background} we review the ABJM background with flavor.  Then, in Section \ref{probes_with_flux}, we consider the embedding of a probe D6-brane with internal flux.  We first present in Section \ref{internal_flux_quantization} the ansatz for the internal components of the worldvolume gauge field that will be used throughout the paper and discuss the corresponding flux quantization condition. In Section \ref{Full_ansatz} we generalize these results to nonvanishing background electric and magnetic fields, as well as to nonzero charge density and currents. We compute the corresponding longitudinal and transverse conductivities in Section \ref{conductivities}.

In Section \ref{em_symmetry} we analyze the residual $SO_+(1,1)$ boost invariance of our system at zero temperature.  An analytic supersymmetric solution of the equations of motion at zero temperature is presented in Section \ref{BPS-sol}.  Section \ref{mesons} is devoted to the analysis of quark-antiquark bound states, \ie, mesons.  In particular, we study the effect of the broken parity on the mass spectrum. In Section \ref{discussion} we summarize our results and discuss possible future directions.

The paper is completed  with several appendices. In Appendix \ref{Background_details} we provide details of our background geometry and discuss the quantization condition of the worldvolume flux obtained by comparison with the ABJ solution. Appendix \ref{EOMs} contains a detailed analysis of the equations of motion of the probe. The kappa symmetry of the embeddings is analyzed in Appendix \ref{kappa}. Finally, the equations governing the fluctuations of the probe are the subject of Appendix \ref{Fluctuations}, where we also estimate the meson masses using a WKB approximation.


\section{The ABJM background with flavor}
\label{Background}
In this section we will review, following \cite{Conde:2011sw, Itsios:2012ev, Jokela:2013qya, Jokela:2012dw}, the background geometry corresponding to the ABJM model with unquenched massless flavors in the smeared approximation. Additional details of this supergravity solution are given in Appendix \ref{Background_details}. The ten-dimensional metric, in string frame,  has the form
\beq
ds^2\,=\,L^2\,\,ds^2_{BH_4}\,+\,ds^2_{6}\,\,,
\label{flavoredBH-metric}
\eeq
where $L$ is the radius of curvature, $ds^2_{BH_4}$ is the metric of a planar black hole in the four-dimensional Anti-de Sitter space, given by
\beq\label{BH4-metric}
 ds^2_{BH_4} = -r^2h(r) dt^2+\frac{dr^2}{r^2h(r)}+r^2\big[dx^2+dy^2\big] \ ,
\eeq
and  $ds^2_{6}$ is the metric of the compact internal six-dimensional manifold. The blackening factor $h(r)$ is given by
\beq
h(r)\,=\,1\,-\,\frac{r_h^3}{r^3} \ ,
\label{blackening-factor}
\eeq
where the horizon radius $r_h$ is related to the temperature $T$ by
$T={3\,r_h\over 4\pi}$. The internal metric $ds^2_{6}$ in (\ref{flavoredBH-metric}) is a deformation of the Fubini-Study metric of ${\mathbb C}{\mathbb P}^3$, realized as an ${\mathbb S}^2$-bundle over ${\mathbb S}^4$. Let 
$ds^2_{{\mathbb S}^4}$ be the standard metric for the unit round four-sphere and let 
$z^i$ ($i=1,2,3$) be  three Cartesian coordinates parameterizing the unit two-sphere ($\sum_i (z^i)^2\,=\,1$). Then, $ds^2_{6}$ can be written as:
\beq
ds^2_{6}\,=\,{L^2\over b^2}\,\,\Big[\,
q\,ds^2_{{\mathbb S}^4}\,+\,\big(d z^i\,+\, \epsilon^{ijk}\,A^j\,z^k\,\big)^2\,\Big] \ ,
\label{internal-metric-flavored}
\eeq
where $A^i$ are the components of the non-Abelian one-form connection corresponding to an $SU(2)$ instanton. In Appendix \ref{Background_details} we give a more explicit representation of the $ds^2_{6}$  line element  in terms of alternative coordinates. 

The parameters $b$ and $q$   in (\ref{internal-metric-flavored})  are constant squashing factors which  encode the effect of the massless flavors in the backreacted metric. Indeed, when $q=b=1$ the metric (\ref{internal-metric-flavored}) is just the canonical Fubini-Study metric of  the ${\mathbb C}{\mathbb P}^3$ manifold with radius $2L$ in  the so-called twistor representation. In this case 
 (\ref{flavoredBH-metric}) is the metric of the unflavored ABJM model at nonzero temperature. When the effect of the delocalized D6-brane sources is taken into account, the resulting metric is deformed as in (\ref{internal-metric-flavored}). It was shown in \cite{Conde:2011sw} that at zero temperature the particular deformation written in (\ref{internal-metric-flavored}) preserves ${\cal N}=1$ SUSY.

 The parameter $b$  in (\ref{internal-metric-flavored}) represents the relative squashing of the ${\mathbb C}{\mathbb P}^3$ part of the metric with respect to the $AdS_4$ part due to the flavor, while $q$ parameterizes an internal deformation which preserves the ${\mathbb S}^4$-${\mathbb S}^2$ split of the twistor representation of ${\mathbb C}{\mathbb P}^3$.  The explicit expressions for the coefficients $q$ and $b$ found in  \cite{Conde:2011sw} are given below. They depend on the number of colors $N$ and flavors $N_f$, as well as on the 't Hooft coupling $\lambda\,=\,N/k$, through the combination
\beq
\hat \epsilon\,\equiv\,{3N_f\over 4k}\,=\,{3\over 4}\,\,{N_f\over N}\,\lambda \ ,
\label{hatepsilon}
\eeq
where the factor $3/4$ is introduced for convenience. It is also useful to define the quantity $\eta$ as:
\beq
\eta\,=\,1\,+\,{3N_f\over 4k}\,=\,1+\hat \epsilon \ ,
\qquad\qquad
\eta\in [1,\infty) \ .
\label{etaNf-k}
\eeq
In terms of the deformation parameter $\hat\epsilon$, the squashing factors $q$ and $b$ are:
\bea
 q & = & 3+\frac{3}{2}\hat\epsilon -2\sqrt{1+\hat\epsilon+\frac{9}{16}\hat\epsilon^2} \ ,   \rc\rc
 b & = & \frac{2q}{q+1}\ .\label{q-b}
\eea
As functions of $\hat \epsilon$, the squashing parameters $q$ and $b$ are monotonically  increasing functions, which approach the values $q\approx 5/3$ and $b\approx 5/4$ as 
$\hat\epsilon\to\infty$. 
Another way to encode the loop effects of the massless sea quarks is to define the screening factor $\sigma$:
\beq
\sigma\,=\,\sqrt{{(4-3b)(2-b)b^3\over 2(b-1)\eta+b}} \ .
\label{screening_sigma}
\eeq
Without flavors, $\sigma = 1$, and as $\hat\epsilon\to\infty$, $\sigma \to 0$.
 The $AdS$ radius $L$ can then expressed in terms of $\lambda$  and the screening factor:
\beq
L^2\,=\,\pi\sqrt{2\lambda}\,\sigma \ .
\label{AdS_radius}
\eeq

The complete solution of type IIA supergravity with sources is endowed with RR two- and four-forms $F_2$ and $F_4$, as well as with a constant dilaton $\phi$ (whose value depends on $N$, $N_f$, and $k$). Their explicit expressions are given in the Appendix \ref{Background_details}.

It is worth mentioning at this point that the background summarized in this section has been generalized in \cite{Bea:2013jxa} to the case in which the quarks are massive (at zero temperature). As we increase the mass of the quarks,  these generalized solutions interpolate between the geometry reviewed above (for zero quark mass) and the unflavored ABJM background (for infinitely massive quarks).


\section{D6-brane probes with flux}
\label{probes_with_flux}

We are interested in dynamics of a massive quark holographically dual to a probe D6-brane with internal flux in the flavored ABJM background. The D6-brane extends along $r$ and the three Minkowski directions and, wraps on the internal manifold a three-cycle topologically equivalent to ${\mathbb R}{\mathbb P}^3={\mathbb S}^3/{\mathbb Z}_2$. This three-cycle will be parameterized by three angles $\alpha$, $\beta$, and $\psi$, and will be characterized by an embedding function $\theta(r)$. With this embedding, the D6-brane then has an induced metric given by (for details see Appendix \ref{Background_details}):
\bear
 {ds^2_{7}\over L^2} &=& r^2\,\left[-h(r)\,dt^2+
 dx^2+dy^2\right]+
  {1\over r^2 }\,
\left({1\over h(r)}+{r^2\,\theta'^{\,2}\over b^2}\,\right)\,dr^2 \rc\rc
&&\qquad +\,{1\over b^2}\,\Big[
q\,d\alpha^2+q\,\sin^2\alpha \,d\beta^2+ \sin^2\,\theta\,\left(\,d\psi\,+\,\cos\alpha\,d\beta\,\right)^2\,\Big] \ ,
\label{induced_metric}
\eear
where $0\le \alpha < \pi$, $0\le \beta, \psi<2\pi$, and $\theta=\theta(r)$ determines the profile of the probe brane. Notice that the second line in (\ref{induced_metric}) is the line element of a squashed ${\mathbb R}{\mathbb P}^3$.

For a supersymmetric configuration at zero temperature, it is possible to use kappa symmetry to find an explicit solution for $\theta(r)$ (see the analysis in \cite{Conde:2011sw} and in Appendix \ref{kappa}).  But, in general we will have to numerically solve the equations of motion to find $\theta(r)$.

The thermodynamic properties of D6-branes embedded in this way were studied in detail in \cite{Jokela:2012dw}. Here we will generalize some of these results by including worldvolume gauge fields.  In particular, we will turn on a nontrivial flux on the internal cycle. In the rest of this section we will determine the form of this internal worldvolume flux which gives rise to a consistent solution of the brane equations of motion.

\subsection{Internal flux}
\label{internal_flux_quantization}

Since we are primarily interested in gapped, QH states, let us focus on Minkowski (MN) embeddings of the probe, in which the brane ends smoothly at a radial position $r_*$ above the horizon, \ie,  $r_* > r_h$.  The D6-brane can cap off smoothly if, at the tip of the brane $r=r_*$, the angle $\theta$ reaches its minimal value $\theta=0$ where an ${\mathbb S}^{1} \subset {\mathbb R}{\mathbb P}^3$ shrinks to zero.  At the tip, the last term of (\ref{induced_metric}) vanishes and the induced metric takes the form:
\beq
\label{induced_metric_tip}
 {ds^2_{7}\over L^2}\Big|_{r=r_*}=r^2\,\big[-h_* dt^2+ dx^2+dy^2\big]+
  {q\over b^2}\,\,
 \Big[d\alpha^2\,+\,\sin^2\alpha \ d\beta^2\Big] \ ,
 \eeq
 where $h_*=h(r=r_*)$. From (\ref{induced_metric_tip}), we see that at the tip of the brane the coordinates $\alpha$ and $\beta$ span a non-collapsing ${\mathbb S}^{2}_{*}$.  As in other probe-brane QH models \cite{Bergman:2010gm, Jokela:2011eb}, we want to turn on a flux of the worldvolume gauge field $F$ on this non-shrinking sphere.

 Of course, this flux must be quantized appropriately. We will adopt the following quantization condition:
\beq
{1\over 2\pi\alpha'}\,\int_{{\mathbb S}^{2}_{*}}\,F\,=\,{2\pi M\over k} \ ,
\qquad\qquad M\in {\mathbb Z}\,\,.
\label{wv_quantization}
\eeq
 Notice that, compared with the ordinary flux quantization condition of the worldvolume gauge field, we are considering in (\ref{wv_quantization}) $M/k$ fractional units of flux. In Appendix \ref{Background_details} we verify that (\ref{wv_quantization}) is the correct prescription for the flux quantization by studying the background without massless flavors, \ie, $N_f = 0$. In this case one can induce an internal $F$ flux through 
${\mathbb S}^{2}_{*}$ by switching on a flat Neveu-Schwarz  $B_2$ field proportional to the K\"ahler form of ${\mathbb C}{\mathbb P}^3$. Then, the quantization condition 
(\ref{wv_quantization}) follows from the fractional holonomy of $B_2$ along the ${\mathbb C}{\mathbb P}^1$ cycle of ${\mathbb C}{\mathbb P}^3$. In this setup the integer $M$ is the number of fractional D2-branes and this configuration is dual to the ABJ model \cite{Aharony:2008gk} with gauge group $U(N+M)_{k}\times U(N)_{-k}$. We also check in Appendix \ref{Background_details} that $M$ can be identified with the Page charge for fractional D2-branes.

Let us now write a concrete ansatz for the internal gauge field $F$. We will represent $F$ in terms of a potential one-form  $A$ given by:
\beq
A\,=\,L^2\,a(r)\,(d\psi+\cos\alpha\,d\beta)\ ,
\label{A_internal}
\eeq
where the $L^2$ factor is introduced for convenience and $a=a(r)$ is a function of the radial coordinate which determines the varying flux on the $(\alpha,\beta)$ two-sphere. The field strength $F=dA$ corresponding to (\ref{A_internal}) is simply:
\beq
F\,=\,L^2\Big[a'(r)\,dr\wedge (d\psi+\cos\alpha d\beta)\,-\,
a(r)\,\sin\alpha\,d\alpha\wedge d\beta\Big] \ ,
\label{F_internal}
\eeq
which restricted to ${\mathbb S}^{2}_{*}$ becomes:
\beq
F\big|_{{\mathbb S}^{2}_{*}}\,=\,-L^2\,a_*\,\sin\alpha\,d\alpha\wedge d\beta \ ,
\eeq
where $a_*\equiv a(r=r_*)$ is the value of the flux function at the tip. It follows that
\beq
\int_{{\mathbb S}^{2}_{*}}\,F\,=\,-4\pi\,L^2\,a_* \ ,
\eeq
and the condition  (\ref{wv_quantization}) quantizes the values of   $a_*$ in the following way:
\beq
a_*\,=\,-{\pi M\over k L^2}\,\,,
\qquad\qquad M\in {\mathbb Z} \ .
\label{a*_quantization}
\eeq
Let us denote the value of the flux function at the tip as:
\beq
a_*\,=\,-Q \ .
\label{a_*_Q}
\eeq
To write the quantization condition (\ref{a*_quantization}) in terms of physical quantities, recall that the $AdS$ radius $L$ can be written as in (\ref{AdS_radius}). Plugging this into (\ref{a*_quantization}), we find the following quantization condition for $Q$:
\beq
Q\,=\,{\sqrt{\lambda}\over \sqrt{2}\,\sigma}\,{M\over N} \ ,
\qquad\qquad M\in{\mathbb Z} \ .
\eeq

Using the ansatz (\ref{F_internal}) for the internal flux, we can try to find a solution for a MN embedding of the probe D6-brane.  In Appendix \ref{EOMs} we check that (\ref{F_internal}), together with embedding ansatz corresponding to the induced metric (\ref{induced_metric}), is a consistent truncation of the equations of motion of the probe. 

At zero temperature, we have found an analytic solution for $\theta(r)$ and the flux function $a=a(r)$ which preserves two of the four supercharges of the ${\cal N}=1$ superconformal background. The explicit calculations are performed in Appendix \ref{kappa} with the use of kappa symmetry. Here we just quote the result for $\theta(r)$ and $a(r)$:
\bear
\cos\theta(r) &=&\Big({r_*\over r}\Big)^{b} 
\label{theta_SUSY_EBzero} \\
a(r) &=&-Q\,(\cos\theta(r))^{{1\over q}}\,=\,-Q\,\Big({r_*\over r}\Big)^{{b\over q}} \ .
\label{a_SUSY_EBzero}
\eear

However, to realize the quantum Hall states we are interested in, we need to generalize our ansatz 
for the gauge field to include electric and magnetic fields, as well as the components dual to the charge density and current.  We analyze this more general set up in the next subsection.

\subsection{Background fields and currents}
\label{Full_ansatz}

If we want a more general ansatz that includes background electric and magnetic fields and the associated charged current, we need to consider other components of the worldvolume gauge field.  In the standard way, a  magnetic field $B$ and an electric field $E$ are added by turning on the radial zero modes of $F_{xy}$ and $F_{0x}$.  The charge density is holographically related to $F_{r0}$, the longitudinal and Hall currents come from $F_{rx}$ and $F_{ry}$.  We therefore take the worldvolume gauge field to have the form:
\beq
A\,=\,L^2\,\Big[\,a_0(r)\,dt\,+(Et+\,a_x(r))\,dx
\,+\,(B\,x+a_y(r))\,dy\,+\,a(r)\,(d\psi+\cos\alpha\,d\beta)\,\Big] \ .
\label{A_full_ansatz}
\eeq
We can continue to use the induced metric ansatz given by (\ref{induced_metric}), characterized by the embedding function $\theta=\theta(r)$. 

Interestingly, due to our choice in (\ref{A_full_ansatz}) of the internal components of the gauge field, the dependence of the action 
 on the internal angles of the ${\mathbb R}{\mathbb P}^3$ cycle factorizes and consequently, we can consistently take the functions $\theta$, $a$, $a_0$, $a_x$, and $a_y$ to depend only on the radial variable. After integrating over the internal angles $\alpha$, $\beta$, and $\psi$, the DBI action of the D6-brane for our ansatz can be written as:
\beq
S_{DBI}\,=\int d^3x\,dt\,{\cal L}_{DBI} \ ,
\eeq
where the DBI Lagrangian density ${\cal L}_{DBI}$ can be compactly written as:
\beq
{\cal L}_{DBI}\,=\,-{8\pi^2\,L^7\,T_{D6}\,e^{-\phi}\over b^4}\,
{\sqrt{(B^2+r^4)h-E^2}\,\sqrt{q^2+b^4\,a^2}\over \sqrt{h}}\,\sqrt{\Delta} \ ,
\eeq
where $T_{D6}$ is the D6-brane tension and the quantity $\Delta$ is defined to be
\bear
\Delta &=& b^4r^2 h a'^{\,2}\,+\,\sin^2\theta\,\Bigg[b^2+r^2 h\,\theta'^{\,2}\, \rc\rc
&&\qquad\qquad\qquad+
{b^2h\over E^2-(B^2+r^4)h}\,\Big[(B a_0'+E a_y')^2\,+\,r^4 (a_0'^{\,2}-h a_x'^{\,2}+ ha_y'^{\,2})\Big]\Bigg]\,\,.\qquad\qquad
\label{Delta_def_general}
\eear
The Wess-Zumino term of the action is:
\beq
S_{WZ}\,=\,T_{D6}\int_{{\cal M}_7}\,\left(\hat C_7\,+\,\hat C_5\wedge F\,+\,{1\over 2}\,
\hat C_3\wedge F\wedge F\,+\,{1\over 6}\hat C_1\wedge F\wedge F\wedge F \right) \ ,
\label{Wess_Zumino}
\eeq
where, $\hat C_7$,  $\hat C_5$,  $\hat C_3$,  and $\hat C_1$ are the pullbacks to the D6-brane of the RR gauge fields.  All of these terms, except for $\hat C_5\wedge F$, give non-vanishing contributions to the equations of motion.\footnote{One subtlety is that when the backreaction of the flavors is included, the RR field strength $F_2$ is not closed, implying that there is no well-defined RR potential $C_1$.  However, the equations of motion derived from (3.17) only contain $F_2$ and therefore can be generalized to the unquenched case; see Appendix \ref{EOMs} for details.}

In the holographic setup, the charge density is encoded in the bulk by the radial electric displacement field $\tilde D(r)$, which is given by the derivative of the DBI Lagrangian density with respect to the radial component of the physical electric field. From the ansatz (\ref{A_full_ansatz}), and taking into account the physical gauge field $A_{phys}=A/(2\pi\alpha')$, we find:
\beq
\tilde D\,=\,{\partial {\cal L}_{DBI}\over \partial A_{0,phys}'}\,=\,{2\pi \alpha'\over L^2}\,
{\partial {\cal L}_{DBI}\over \partial a_{0}'} \ .
\label{tilde_D_def}
\eeq
We will set $\alpha'=1$ from now on. In order to write  $\tilde D(r)$ in a compact fashion, let us define a function $g(r)$ as:
\beq
g(r)\,=\,{q+\eta\over 2b(2-q)}\,{r^4\,h^{{3\over 2}}\,\sin^2\theta\,\sqrt{q^2+b^4\,a^2}
\over \sqrt{(B^2+r^4)\,h\,-E^2}\,\sqrt{\Delta}} \ .
\eeq
Then, one can show that:
\beq
\tilde D(r)\,=\,{N\sigma^2\over 4\pi}\,\tilde d(r)\ ,
\eeq
where $\sigma$ is the screening factor defined in (\ref{screening_sigma}) and $\tilde d(r)$ is the function:
\beq
\tilde d(r)\,\equiv \,{g\over h}\,\left[\left(1+{B^2\over r^4}\right)\,a_0'\,+\, {BE\over r^4}\,a_y'\,\right]\,\,.
\label{tilded_def}
\eeq
The total charge density is obtained by taking the boundary value of $\tilde D(r)$, which is proportional to:
\beq
d\,=\,\lim_{r\to\infty}\,\tilde d(r) \ .
\eeq
Similarly, the physical currents along the $x$ and $y$ directions are given by:
\beq
J_x\,=\,{2\pi \alpha'\over L^2}\,
{\partial {\cal L}_{DBI}\over \partial a_{x}'} \ ,
\qquad\qquad
\tilde J_y\,=\,{2\pi \alpha'\over L^2}\,
{\partial {\cal L}_{DBI}\over \partial a_{y}'} \ .
\label{Jxy_def}
\eeq
One can readily prove that:
\beq
J_x\,=\,{N\sigma^2\over 4\pi}\,j_x\,\,,
\qquad\qquad\qquad
\tilde J_y\,=\,{N\sigma^2\over 4\pi}\,\tilde j_y \ ,
\eeq
where $j_x$ turns out to be:
\beq
j_x\,=\,- g\,a_x' \ ,
\label{eom_ax_jx}
\eeq
and $\tilde j_y(r)$ is:
\beq
\tilde j_y(r)\,\equiv \,g\,\left[-\left(1-{E^2\over r^4 h}\right)\,a_y'\,+\,
{BE\over r^4 h}\,a_0'\,\right] \ .
\label{tildejy_def}
\eeq

The equations of motion for the probe are worked out in detail in Appendix \ref{EOMs}.
In particular, $J_x$ is constant in $r$  (see  (\ref{eom_ax_general_case})) and represents the longitudinal current parallel to the electric field. 
On the other hand, $\tilde J_y(r)$ depends on the holographic variable. The transverse current $J_y$ is obtained as the value of $\tilde J_y(r)$  at the UV boundary $r\to\infty$ which, according to (\ref{Jxy_def}),  is determined from the limit:
\beq
 j_y\,=\,\lim_{r\to\infty}\,\tilde j_y(r) \ .
\eeq

The radial dependence of $\tilde d$ and $\tilde j_y$ is determined by the $a_0$ and $a_y$ equations of motion, (\ref{eom_a0_general_case}) and (\ref{eom_ax2_general_case}). With the definitions introduced above, they can be simply written as:
\bear
\label{eom_dtilde}
\partial_r\,\tilde d &=& B(\eta\,\cos\theta\,a'-a\,\sin\theta\,\theta') \\
\label{eom_jytilde}
\partial_r\,\tilde  j_y &=& E(\eta\,\cos\theta\,a'-a\,\sin\theta\,\theta')\,\,.
\eear
In the unflavored case $\eta=1$, these two equations (\ref{eom_dtilde}) and (\ref{eom_jytilde}) can be integrated once because their right-hand side is proportional to $\partial_r (a\cos\theta)$. Indeed, for the unflavored background $a_0(r)$ and $a_y(r)$ are cyclic and can be eliminated by performing the appropriate Legendre transformation. 

This is not the case, however, when $\eta\not=1$. We can formally integrate (\ref{eom_dtilde}) and (\ref{eom_jytilde}), defining the integral $I(r)$ as:
\bear
\label{I_integral_def}
I(r) &\equiv& \,\int_r^{\infty}\,\Big(\eta\,\cos\theta(\bar r)\,a'(\bar r)\,-\,a(\bar r)\,\sin\theta(\bar r)\,\theta'(\bar r)\Big)\,d\bar r  \\
\label{I_partial_integral}
&=& \,-\cos\theta(r)\,a(r)\,+\,(\eta-1)\,\int_r^{\infty} \cos \theta(\bar r)\,a'(\bar r)\,d\bar r  \ ,
\eear
where we have integrated by parts to obtain the second line.  Clearly, 
\beq
\lim_{r\to \infty}\,I(r)\,=\,0 \ ,
\eeq
and equations (\ref{eom_dtilde}) and (\ref{eom_jytilde}) can be written as
\beq
\tilde d(r)\,=\,d\,-\,B\,I(r)\ ,
\qquad\qquad
\tilde j_y(r)\,=\, j_y\,-\,E\,I(r)\ .
\label{tilde_d_jy_integrated}
\eeq

Since $a_0$ and $a_y$ are no longer cyclic, we need a new strategy to solve the equations of motion. Interestingly, there is still one conserved quantity associated with the equations of motion for $A_0$ and $A_y$.  Eq.~({\ref{constant_of_motion}) can be recast as the radial independence of the quantity:
\beq
\Pi \,\equiv\,E\,\tilde d(r)\,-\,B\,\tilde j_y(r)\ .
\label{Pi_def}
\eeq
Indeed, the equation $\partial_r\,\Pi=0$ follows immediately from (\ref{eom_dtilde}) and (\ref{eom_jytilde}). This implies that $\Pi$ can be written in terms of quantities evaluated at the boundary:
\beq
\Pi\, = \, E \, d\,-\,B\, j_y\ .
\label{Pi_UV}
\eeq

One can now try to write $a_0'$, $a_y'$, and $a_x'$ in terms of the embedding function $\theta(r)$ and the flux function $a(r)$. Let us work this out in detail. First, we notice that one can invert Eqs.~(\ref{tilded_def}) and (\ref{tildejy_def}) and write $a_0'$ and $a_y'$ in terms of $\tilde d$ and $\tilde j_y$:
\beq
a_0'\,=\,{h\Big(1-{E^2\over r^4\,h}\Big)\,\tilde d\,+\,{EB\over r^4}\,\tilde j_y\over
g\Big(1+{B^2\over r^4}\,-\,{E^2\over r^4\,h}\Big)}\ ,
\qquad\qquad
a_y'\,=\,{{EB\over r^4}\,\tilde d\,-\,\Big(1+{B^2\over r^4}\Big)\,\tilde j_y
\over
g\Big(1+{B^2\over r^4}\,-\,{E^2\over r^4\,h}\Big)}\ .
\label{a_0_y_disp_quenched}
\eeq
Notice that (\ref{a_0_y_disp_quenched}) are not actually solutions for $a_0'$ and $a_y'$ since $g$ on the right-hand side is written in terms of these same fields. However, one can write an expression of $g$ in terms of $\theta$ and $a$. Let us define $X$ as:
\beq
X\equiv h\left(1+{B^2\over r^4}\,-\,{E^2\over r^4\,h}\right)
\Bigg[\left({q+\eta\over 2 b^2 (2-q)}\right)^2
r^4 h (q^2+b^4\,a^2)\,\sin^2\theta +h\tilde d^{\,2}-j_x^2-\tilde j_y^{\,2}\Bigg]- \,
{\left(hB\tilde d-E\tilde j_y\right)^2\over r^4}\ .
\eeq
Then, after some calculation, we obtain:
\beq
\label{gdef}
g\,=\,{\sin\theta\,\sqrt{X}\over
\Big(1+{B^2\over r^4}\,-\,{E^2\over r^4\,h}\Big)\,
\sqrt{b^2\,r^2\,h\,a'^{\,2}\,+\,\sin^2\theta\,(1+{r^2\over b^2}\,h\,\theta'^{\,2})}}\,\,.
\eeq
Therefore, we have for $a_0'$, $a_x'$, and $a_y'$:
\bear
&&a_0'\,=\,{
\sqrt{b^2\,r^2\,h\,a'^{\,2}\,+\,\sin^2\theta\,(1+{r^2\over b^2}\,h\,\theta'^{\,2})}\over \sin\theta\sqrt{X}}\,
\Bigg[h\left(1-{E^2\over r^4\,h}\right)\,\tilde d\,+\,{EB\over r^4}\,\tilde j_y\Bigg]\,\,,\rc\rc
&&a_x'\,=\,-{
\sqrt{b^2\,r^2\,h\,a'^{\,2}\,+\,\sin^2\theta\,(1+{r^2\over b^2}\,h\,\theta'^{\,2})}\over 
\sin\theta\sqrt{X}}\,
\left(1+{B^2\over r^4}\,-\,{E^2\over r^4\,h}\right)\,j_x\,\,,\rc\rc
&&a_y'\,=\,{
\sqrt{b^2\,r^2\,h\,a'^{\,2}\,+\,\sin^2\theta\,(1+{r^2\over b^2}\,h\,\theta'^{\,2})}\over
 \sin\theta\sqrt{X}}\,
\Bigg[
{EB\over r^4}\,\tilde d\,-\,\left(1+{B^2\over r^4}\right)\,\tilde j_y
\Bigg]\,\,.
\label{aprimes_tilded_jy}
\eear
The right-hand sides of (\ref{aprimes_tilded_jy}) contain the radial functions $\tilde d$ and $\tilde j_y$, which in turn can be written as in (\ref{tilde_d_jy_integrated}) in terms of the constants $d$ and $j_y$, and the integral $I(r)$ defined in (\ref{I_integral_def}). 

In principle, we could  use (\ref{aprimes_tilded_jy}) to eliminate $a_0'$, $a_x'$, and $a_y'$ from the equations of motion and to reduce the system to an effective problem for the functions 
$\theta(r)$ and $a(r)$. However, when $\eta\not=1$, the functions $\tilde d(r)$ and $\tilde j_y$ depend non-locally on $\theta(r)$ and $a(r)$ and the corresponding reduced equations of motion would be a system of integro-differential equations for $\theta(r)$ and $a(r)$, which does not seem to be very easy to solve in practice.  In the case in which there are no flavors backreacting on the geometry, \ie, when $\eta=q=b=1$, the integral $I(r)$ is just 
$I=-\cos\theta\, a$ and we can write $\tilde d$ and $\tilde j_y$ simply as
$\tilde d\,=\,d\,+\,B\cos\theta\,a$ and 
$\tilde j_y\,=\,j_y+E\cos\theta \,a$. Thus, 
in this quenched case one can eliminate the gauge fields $a_0$, $a_x$, and $a_y$ and reduce the problem to a system of two coupled, second-order differential equations for  $\theta(r)$ and $a(r)$.

\subsection{Minkowski embeddings}

Having obtained the equations of motion for the D6-brane probe, the next step is to try to solve them.  Although, as we will discuss in Section \ref{BPS-sol}, there are special analytic BPS solutions, in general we will have to resort to numerics. 

Probe brane solutions are categorized into two classes by their IR behavior.  The generic solution is a black hole embedding, in which the brane falls into the horizon; these correspond holographically to gapless, compressible states.  In certain special circumstances, the brane can end smoothly at some $r=r_*$ when a wrapped cycle shrinks to zero size; these are Minkowski (MN) embeddings.  MN solutions with broken parity correspond to gapped, quantum Hall states.

As discussed above in Section \ref{internal_flux_quantization}, for a D6-brane probe in the flavored ABJM background, MN embeddings occur when $\theta(r_*) = 0$ for some $r_*$.  In order to have a physical, finite-energy solution, the embedding $\theta(r)$ and the worldvolume gauge field $F$ must be regular at the tip; that is, the induced metric (\ref{induced_metric}) must be smooth, and $a'$, $a_0'$, $a_x'$, $a_y'$ must all be finite. Given that the function $g$ (\ref{gdef}) vanishes at the tip of the brane, the regularity of $a_0'$ and $a_y'$ at the tip, combined with (\ref{tilded_def}) and (\ref{tildejy_def}), implies that
\beq
\label{tip_conditions}
\tilde d(r_*)\,=\, j_x \, = \,\tilde j_y(r_*)\,=\,0\,\,.
\eeq
We can interpret this condition to mean that there are no sources at the tip, which is physically sensible as the D6-brane could not support such a source.  Suppose that $\tilde d(r_*) \not= 0$; this radial displacement field would have to be sourced, for example, by fundamental strings stretching from the horizon.  Due to the shrinking cycle, the effective radial tension of the D6-brane vanishes at the tip, so these strings would then pull the D6-brane into the horizon, resulting in a black hole embedding.

The filling fraction $\nu$ is defined by
\beq
\nu \, = \, 2\pi \, \frac{D_{phys}}{B_{phys}} \ ,
\eeq
where the physical magnetic field $B_{phys}$ is related to $B$ by
\beq
B_{phys}\,=\,{L^2\over 2\pi}\,B\,=\,\sqrt{{\lambda\over 2}}\,\sigma\,B\ .
\eeq
Combining (\ref{tip_conditions}) with (\ref{tilde_d_jy_integrated}) gives $d = BI(r_*)$, and the filling fraction for MN solutions is therefore
\beq
\label{nu_MN}
\nu\,=\,{N\sigma\over \sqrt{2\lambda}}\,\frac{d}{B}\,=\,{N\sigma\over \sqrt{2\lambda}}\,I(r_*) \ ,
\eeq
or, more explicitly using (\ref{I_partial_integral}),
\beq
\nu\,=\,{M\over 2}\,\left[
1\,+\,(\eta-1)\,\int_{r_*}^{\infty}\,
\cos\theta(r)\,{a'(r)\over Q}\,dr\,\right]\ ,
\label{nu_flux}
\eeq
where $M$ is the quantization integer and $Q$ is minus the flow function at the tip (see (\ref{a_*_Q})).  Note that,  (\ref{nu_flux}) shows explicitly that, for a QH state with nonzero charge density, a nonzero flux is required. Moreover, $\nu$ is the sum of two contributions. The first term in (\ref{nu_flux}) is proportional to the flux at the tip. The second term is only nonzero in the unquenched case $\eta\not=1$ and contains an integral from the tip to the boundary. In terms of $N_f$ and $k$, $\nu$ takes the form:
\beq
\nu\,=\,{M\over 2}\,\left[1\,+\,{3N_f\over 4 k}\,
\int_{r_*}^{\infty}\,
\cos\theta(r)\,{a'(r)\over Q}\,dr\,\right]\ .
\label{filling_flux}
\eeq
It follows that $\nu$ is a half-integer in the quenched case but gets corrections due to the massless sea quark loops in the unquenched Veneziano limit.  

Numerically integrating the equations of motion, we have verified that there are MN solutions obeying the tip regularity conditions (\ref{tip_conditions}).  At this point, we will be content with evidence for MN solutions with nonzero charge density $d$ and magnetic field $B$.  We will defer a more thorough study of the possible MN solutions to the future.

\section{Conductivities}
\label{conductivities}

We are interested in analyzing the longitudinal and transverse conductivity of our configurations. In order to relate these quantities with the variables we have employed, let us point out that the physical electric field $E_{phys}$ is related to the quantity $E$  used above  as:
\beq
E_{phys}\,=\,{L^2\over 2\pi}\,E\,=\,\sqrt{{\lambda\over 2}}\,\sigma\,E\ .
\eeq
The longitudinal  and transverse conductivities $\sigma_{xx}$ and $\sigma_{xy}$  are defined in terms of $J_x$, $J_y\equiv\tilde J_y(r\to \infty)$ and  $E_{phys}$ as:
\beq
\sigma_{xx}\,=\,{J_x\over E_{phys}}\ ,
\qquad\qquad
\sigma_{xy}\,=\,{J_y\over  E_{phys}}\ .
\eeq
The conductivities can be written as
\beq
\sigma_{xx}\,=\,{N\sigma\over 2\pi\sqrt{2\lambda}}\,{j_x\over E}\ ,
\qquad\qquad
\sigma_{xy}\,=\,{N\sigma\over 2\pi\sqrt{2\lambda}}\,{j_y\over E}\ .
\eeq
In the next two subsections we obtain formulas for $\sigma_{xx}$ and $\sigma_{xy}$ for the two types of embeddings (Minkowski and black hole) of the D6-brane probe.

\subsection{Quantum Hall states}
\label{QH_conductivity}

Let us now suppose that we have a Minkowski (MN) embedding. To compute the conductivities, we will adapt the method of \cite{Bergman:2010gm, Jokela:2011eb} but with some new twists.  In particular, we use the invariance of $\Pi$ under the the holographic flow.  The conductivity comes directly from the condition (\ref{tip_conditions}) that there are no charge sources at the tip $r=r_*$. Since $j_x$ in (\ref{eom_ax_jx}) must be equal to zero, 
\beq
\sigma_{xx}\,=\,0 \ .
\eeq
Furthermore, (\ref{tip_conditions}) implies that $\Pi$ vanishes at $r=r_*$ and, since it is radially invariant, $\Pi=0$  at all values of $r$. From (\ref{Pi_UV}), we see that this is equivalent to $E\, d = B\, j_y$; the Hall conductivity is then:
\beq
\sigma_{xy}\,=\,{N\sigma\over 2\pi\sqrt{2\lambda}}\,{j_y\over E}\,
=\,{N\sigma\over 2\pi\sqrt{2\lambda}}\,{d\over B}\ .
\eeq
From (\ref{nu_MN}), we find that
\beq
\sigma_{xy}\,=\,{\nu\over 2\pi} \ , 
\label{sigma_xy_flux}
\eeq
which is exactly what one would expect for a QH state.

\subsection{Gapless states}

Let us now consider black hole embeddings, in which the D6-brane crosses the horizon at $r=r_h$. These embeddings correspond to gapless states. To compute the conductivity, we employ the pseudohorizon argument of \cite{Karch:2007pd} to Eq.~(\ref{aprimes_tilded_jy}).  Let $r=r_p$ be the position of the pseudohorizon, which is determined by the conditions:
\bear
h_p\,(r_p^4+B^2) &=& E^2 \rc\rc
j_x^2\,+\,\tilde j_y^2(r_p) &=& \left({q+\eta\over 2 b^2 (2-q)}\right)^2\,h_p\,r_p^4\,\sin^2\theta_p\,\left(q^2+b^4\,a_p^2\right)\,+\,h_p\,\tilde d(r_p)^2 \rc\rc
E\,\tilde j_y(r_p) &=& B\,h_p\,\tilde d(r_p) \ ,
\eear
where $h_p\equiv h(r_p)$, $\theta_p\equiv \theta(r_p)$, and $a_p\equiv a(r_p)$. 
It follows that the currents in $x$- and $y$-directions are given by:
\bear
j_x&=&\sqrt{h_p}\,
\Bigg[\left(1-{B^2 h_p\over E^2}\right)\,\tilde d^{\,2}(r_p)\,+\,
\left({q+\eta\over 2 b^2 (2-q)}\right)^2\,r_p^4\,(q^2+b^4\,a_p^2)\,\sin^2\theta_p
\Bigg]^{{1\over 2}} \rc\rc
j_y&=&{B\,h_p\over E}\,d\,+\,E\,\left[1-{B^2 h_p\over E^2}\right]\,I(r_p) \ .
\eear
Notice that the previous expression involves the value of the integral $I$ extended between the $r_p$ and the boundary. Therefore,  the conductivities are:
\bear
\sigma_{xx}&=&{N\sigma\over 2\pi\sqrt{2\lambda}}\,
{\sqrt{h_p}\over E}\,
\Bigg[\left(1-{B^2 h_p\over E^2}\right)\,\tilde d^{\,2}(r_p)\,+\,
\left({q+\eta\over 2 b^2 (2-q)}\right)^2\,r_p^4\,(q^2+b^4\,a_p^2)\,\sin^2\theta_p
\Bigg]^{{1\over 2}} \rc\rc
\sigma_{xy}&=&{N\sigma\over 2\pi\sqrt{2\lambda}}\,\Bigg[
{B\,h_p\over E^2}\,d\,+\,\left[1-{B^2 h_p\over E^2}\right]\,I(r_p)\Bigg]
\ .
\label{metallic_conductivities}
\eear
For small electric field, $r_p$ is close to the horizon radius $r_h$. At first order in $E^2$ we can estimate $r_p$ as:
\beq
r_p\,\approx \,r_h\,\left(1\,+\,{E^2\over 3(r_h^4+B^2)}\right) \ .
\eeq
With this result we can write $h_p$ approximately as:
\beq
h_p\approx {E^2\over r_h^4+B^2}\ .
\eeq
Applying these results to (\ref{metallic_conductivities}), we obtain the linear conductivities:
\bear
 \label{longitudentalconductivity}
\sigma_{xx} &\approx& {N\sigma\over 2\pi\sqrt{2\lambda}}\,
{r_h^2\over  r_h^4+B^2}\,
\Bigg[\tilde d_h^2\,+\,\left({q+\eta\over 2 b^2 (2-q)}\right)^2\,
(r_h^4+B^2)\,(q^2+b^4\,a_h^2)\,\sin^2\theta_h\Bigg]^{{1\over 2}} \\
\label{Hallconductivity}
\sigma_{xy} &\approx& {N\sigma\over 2\pi\sqrt{2\lambda}}\,\Bigg[
{B\,\tilde d_h\over r_h^4+B^2}\,+\, I_h\Bigg] \ ,
\eear
where $I_h\equiv I(r_h)$ is defined in (\ref{I_partial_integral}).

These conductivities are analogous to the conductivities found in the metallic phases of other similar probe brane models, for example \cite{Lifschytz:2009si, Bergman:2010gm, Jokela:2011eb}.  One important difference is that here, the unquenched sea quarks reduce the conductivity by the screening factor $\sigma$.

The longitudinal conductivity (\ref{longitudentalconductivity}) receives contributions from two sources:  the first term under the square root is due to the charge density at the horizon $\tilde d_h$, and the other term can be interpreted as being due to thermal pair production.  At vanishing magnetic field and nonzero charge density, $\sigma_{xx}$ diverges as $r_h^{-2}$.  Charge carriers can only scatter off the thermal bath, and at zero temperature, momentum conservation implies an infinite DC conductivity.  For nonzero $B$, $\sigma_{xx}$ vanishes in the zero-temperature limit, as implied by Lorentz invariance.

The Hall conductivity  (\ref{Hallconductivity}) is the sum of two terms, which appear to correspond to the contributions of two types of charge carriers:  the charges at the horizon $\tilde d_h$, which are sensitive to the heat bath and contribute to $\sigma_{xx}$, and the charges $B I_h = d-\tilde d_h$, which are smeared radially along the D6-brane and do not interact with the dissipative heat bath at all.  In the limit where $d/B \to I_h$, \ie, $\tilde d \to 0$, the Hall conductivity smoothly approach the results found above for the MN embedding (\ref{sigma_xy_flux}). Varying the $d/B$ from zero to $I_h$ and plotting the conductivity on the $(\sigma_{xy},\sigma_{xx})$-plane is expected to reproduce the behavior, seen also in \cite{Jokela:2011eb}, qualitatively similar to the semi-circle law experimentally observed in QH systems \cite{Dykhne}.

\section{Boost invariance at zero temperature}
\label{em_symmetry}

At zero temperature and before adding an electric field, the system is Lorentz invariant.  
In probe brane constructions, the zero-temperature limit of a black hole embedding is often problematic.  However, Minkowski embeddings are perfectly well defined in the zero-temperature limit since the brane never reaches the black hole horizon.  One important feature of this model and others in its class \cite{Bergman:2010gm, Jokela:2011eb, Kristjansen:2012ny}, is that MN embeddings can occur at nonzero charge density.

Turning on an electric field in the $x$-direction breaks rotation invariance, and the full Lorentz symmetry is reduced to a (1+1)-dimensional subgroup: boosts in the $y$-direction form a set of $SO_+(1,1)$ transformations which rotate the electromagnetic field and the currents. When $h=1$, the equations of motion studied in Section \ref{Full_ansatz}  and Appendix \ref{EOMs} are not modified under these transformations.

In terms of the boundary variables, a boost with rapidity $\gamma$ acts as 
\beq
\begin{pmatrix}
E\\  B 
\end{pmatrix}
\to\,{\cal M}_{\gamma}\,\begin{pmatrix}
E\\  B 
\end{pmatrix}
\,\,,\qquad\qquad
\begin{pmatrix}
d\\  j_y
\end{pmatrix}
\to\,{\cal M}_{\gamma}\,\begin{pmatrix}
d\\  j_y
\end{pmatrix}\ .
\label{proper_o11}
\eeq
where ${\cal M}_{\gamma}$ is the symmetric matrix:
\beq
{\cal M}_{\gamma}\equiv 
\begin{pmatrix}
  \cosh \gamma &&\sinh \gamma \\
  {} && {} \\
 \sinh \gamma && \cosh \gamma
 \end{pmatrix}\ .
 \label{M_alpha}
\eeq
The transverse conductivity $\sigma_{xy}$ is invariant under the boost because it is determined only by the flux (\ref{sigma_xy_flux}). 

The boundary electromagnetic fields and currents are packaged holographically in the bulk worldvolume field strength $F$. From the transformation properties of $F$ due to a boost in the bulk, one can reproduce the transformation (\ref{proper_o11}) of $E$ and $B$ and see that the radial components of $F$ transform as
\beq
\begin{pmatrix}
F_{r0}\\  F_{ry}
\end{pmatrix}
=
\begin{pmatrix}
a_0'\\  a_y'
\end{pmatrix}
\to\,{\cal M}_{-\gamma}\,\begin{pmatrix}
a_0'\\ a_y'
\end{pmatrix}\ ;
\eeq
therefore, the symmetry acts contravariantly on $(a_0',  a_y')$. Using Eqs.~(\ref{tilded_def}) and (\ref{tildejy_def}), one can demonstrate that the functions $\tilde d$ and $\tilde j_y$ also rotate via ${\cal M}_{\gamma}$, namely:
\beq
\begin{pmatrix}
\tilde d\\  \tilde j_y
\end{pmatrix}
\to\,{\cal M}_{\gamma}\,\begin{pmatrix}
\tilde d\\  \tilde j_y
\end{pmatrix}
\eeq
which matches, for $r = \infty$, the transformation (\ref{proper_o11}) of $d$ and $j_y$. One can also check that  the quantity $\Pi$ defined in (\ref{Pi_def}) is invariant.

Apart from the boosts, the equations of motion are also invariant under the two types of discrete operations, which are the elements of $O(1,1)$ not connected to the identity. The first of these operations is the 
electric field inversion ${\cal P}_{E}$, which acts as:
\beq
{\cal P}_{E}: \ 
\begin{pmatrix}
E\\  B 
\end{pmatrix}
\to\,
 \begin{pmatrix}
-E\\  \ B 
\end{pmatrix}\  , \ \ 
\begin{pmatrix}
a_0'\\  a_y'
\end{pmatrix}\to
\begin{pmatrix}
\ a_0'\\  -a_y'
\end{pmatrix}\ , \ \
\begin{pmatrix}
\tilde d\\  \tilde j_y
\end{pmatrix}\,\to\,
\begin{pmatrix}
\ \tilde d\\ - \tilde j_y
\end{pmatrix}\ .
\label{PE_def}
\eeq
 Under ${\cal P}_E$, the function $\Pi$ changes its sign, \ie, $\Pi$ behaves as a pseudoscalar.  However, the conductivity $\sigma_{xy}$ is left invariant.
 Similarly, the equations of motion are invariant under a magnetic field inversion 
 ${\cal P}_{B}$, defined as:
\beq
{\cal P}_{B}:\
\begin{pmatrix}
E\\  B 
\end{pmatrix}
\to\,
 \begin{pmatrix}
\ E\\  -B 
\end{pmatrix}\ , \ \
\begin{pmatrix}
a_0'\\  a_y'
\end{pmatrix}\to
\begin{pmatrix}
-a_0'\\  \,\,\,a_y'
\end{pmatrix}\ , \ \
\begin{pmatrix}
\tilde d\\  \tilde j_y
\end{pmatrix}\,\to\,
\begin{pmatrix}
-\tilde d\\ \ \ \tilde j_y
\end{pmatrix}\ .
\eeq
Under this transformation, $\Pi$ again changes its sign and $\sigma_{xy}$ is again invariant. 

We can use the $O(1,1)$ symmetry to classify the different configurations in terms of the sign of the following quadratic forms 
\beq
{\cal Q}_1\equiv E^2\,-\,B^2 \ ,
\qquad\qquad
{\cal Q}_2\equiv d^{\,2}\,-\,j_y^{\,2}\ ,
\qquad\qquad
{\cal Q}_3\,\equiv B\,d\,-\,E\,j_y \ ,
\eeq
which are  left invariant by the $O(1,1)$ transformations.  We will call solutions with ${\cal Q}_1 >0 $ ``electric-like'' , and those with ${\cal Q}_1 <0$ ``magnetic-like''.  We can also have solutions with  ${\cal Q}_1=0$, which we call ``null" solutions. Notice that an electric-like (magnetic-like) solution can be connected continuously to a solution with $B=0$ ($E=0$) and, an electric-like solution cannot be transformed into a magnetic-like one.  We could similarly classify the solutions according to the sign of ${\cal Q}_2$ and  ${\cal Q}_3$. 

For MN embeddings, however, ${\cal Q}_2$ and ${\cal Q}_3$ are not independent but are rather proportional to ${\cal Q}_1$.  The regularity at the tip (\ref{tip_conditions}) implies $d/B = j_y/E = I(r_*)$.  From these relations, we find
\beq
{\cal Q}_2 = {\cal Q}_3 I(r_*) =  - {\cal Q}_1 I(r_*)^2 \ .
\eeq

In the next section we will find an analytic solution to the equations of motion for which the three ${\cal Q}_i$ invariants vanish. Intuitively, one would think that these null solutions have a large degree of symmetry. In particular, they are related to the $E=B=a_0'=a_y'=0$ solution by the infinite boost ${\cal M}_{-\infty}$.  Indeed, we will prove that these are BPS solutions preserving a fraction of the supersymmetry of the background.

\section{The BPS solution}
\label{BPS-sol}

In this section we find a simple, exact MN solution of the zero-temperature equations of motion (\ref{eom_a0_general_case})-(\ref{eom_theta_general_case}).  This solution preserves one supercharge, or one quarter of the supersymmetry of the background.  Accordingly, we will refer to this solution as the BPS solution.

Let us first consider the probe D6-brane in the absence of electric and magnetic fields and with $a_0' = a_y' = 0$.  At zero temperature, we found a SUSY solution in Section \ref{internal_flux_quantization} for which the embedding function $\theta(r)$ and the flux function $a(r)$ satisfy the system of first-order BPS equations (\ref{BPSeq_theta}) and (\ref{BPSeq_a}) derived in Appendix \ref{kappa}:
\beq
r\,\theta'\,=\,b\,\cot\theta\ ,
\qquad\qquad\qquad
{a'\over a}\,=\,-{b\over q\,r}\ .
\label{BPSeq_theta_a}
\eeq

We now generalize this supersymmetric solution to include electric and magnetic fields, as well as charge density and current, provided they satisfy a BPS condition:
\beq
E\,=\,B \ ,
\qquad\qquad
a_0'\,=\,-a_y' \ .
\label{EB_BPS}
\eeq
In addition, we take $a_x'=0$. 
Notice that, since $h=1$, (\ref{EB_BPS}) implies (\ref{constant_of_motion}) is trivially satisfied with the constant on the right-hand side equal to zero.  
Moreover, the equations of motion for $a_0$ (\ref{eom_a0_general_case}) and $a_y$  (\ref{eom_ax2_general_case}) become equivalent. The quantity $\Delta$ defined in (\ref{Delta_def_general}) greatly simplifies and satisfies:
\beq
{\sqrt{q^2+b^4\,a^2}\over \sqrt{\Delta}}\Bigg|_{BPS}\,=\,{q\over b} \ .
\eeq
As we saw in Section \ref{QH_conductivity}, for MN embeddings $\Pi = 0$.  Combining this fact with (\ref{EB_BPS}), yields $\tilde d(r)=\tilde j_y(r)$, and in particular, $d=j_y$. 

Using these results it is straightforward to verify that the equations (\ref{eom_theta_general_case}) and (\ref{eom_apsi_general_case}) for 
$\theta(r)$ and $a(r)$ become:
\bear
&&\partial_r\,\big[r^4\,\sin^2\theta\,\theta'\big]\,-\,b^2\,r^2\,  
\left( 1+\frac{b^4}{q^2}a^2-b^2 r^2 a'^2  \right) 
 \cot\theta\,-(3-2b)\,b\, 
r^2\sin\theta\cos\theta\,=\,0\ ,\rc\rc
&&\partial_r\,\big[r^4\,a'\big]\,+\,{b\over q}\,\Big(3-{b\over q}\Big)\,r^2\,a\,=\,0 \ .
\label{eom_theta_a}
\eear
Note that these equations are just the same as in the $E=B=a_0'=a_y'=0$ case. One can also check that the second-order equations (\ref{eom_theta_a}) are satisfied if the first-order ones in (\ref{BPSeq_theta_a}) are fulfilled.  So, the solutions for $\theta(r)$ and $a(r)$ are just as in Section \ref{internal_flux_quantization} (see (\ref{theta_SUSY_EBzero}) and (\ref{a_SUSY_EBzero})):
\bear
\cos\theta &=& \left({r_*\over r}\right)^{b} 
\label{theta_SUSY} \\ 
a &=& -Q\,\Big({r_*\over r}\Big)^{{b\over q}}\,=\,-Q\,(\cos\theta)^{{1\over q}} \ .
\label{a_SUSY}
\eear

It remains to solve for $a_0$. Its equation of motion (\ref{eom_a0_general_case}) simplifies to:
\beq
\partial_r\,\left[r^2\,\sin^2\theta\,a_0'\right]\,=\,-2B\,{(2-q)\,b^3\over q^2}\,
{a\cos\theta\over r} \ .
\label{eom_a0_BPS}
\eeq
Plugging in the solutions (\ref{theta_SUSY}) for $\theta(r)$ and (\ref{a_SUSY}) for $a(r)$, it is now straightforward to integrate (\ref{eom_a0_BPS}). Using the relation $q=b/(2-b)$, we get:
\beq
a_0'\,=\,{1\over r^2}\,\,{1\over 1-\Big({r_*\over r}\Big)^{2b}}\,\Bigg[\frac{2b^2(2-q)}{q(q+\eta)} d\,-\,
(4-3b)(2-b)b\,Q\,B\,\Big({r_*\over r}\Big)^{2}\Bigg] \ ,
\eeq
The regularity condition of $a_0'$ at the tip of the brane fixes a relation between $d$, $Q$, and $B$:
\beq
d \,=\,\frac{(q+\eta)(2-b)}{2}\,Q\,B \ .
\label{a_0_infty}
\eeq
From the first equality of (\ref{nu_MN}), the filling fraction $\nu$ for this SUSY solution is then:
\beq
\label{nu_SUSY_a}
\nu\,=\,{N\sigma\over \sqrt{2\lambda}}\,\frac{d}{B}\,= (q+\eta)(2-b) \, \frac{M}{4} \  .
\eeq
As we found in (\ref{nu_flux}) for general MN solutions, the filling fraction is proportional to the internal flux.
In addition, (\ref{nu_SUSY_a}) can be rewritten as
\beq
\nu\,=\,\left[1+{3N_f\over 8k}\,\big(1-\gamma_{m}\big)\,\right]\,{M\over 2} \ ,
\label{nu_SUSY_b}
\eeq
where $\gamma_m\,=\,b-1$ is anomalous dimension of the quark mass (see \cite{Conde:2011sw, Jokela:2013qya}). Notice that $\gamma_m$ depends on $N_f/k$ and controls the  coefficient of the contribution of the quarks loops to $\nu$. 

For the BPS solution, the integral $I(r)$ can be explicitly performed using the expressions (\ref{theta_SUSY}) and (\ref{a_SUSY}) for $\theta(r)$ and $a(r)$.  In particular,
\beq
I(r_*) = Q + (\eta - 1) \int_{r_*}^{\infty}\,\cos\theta a'\,dr\,=\, {q+\eta \over q+1} \, Q \ .
\eeq
As a cross-check, we can compute the filling fraction using the second equality in (\ref{nu_MN}):
\beq
\nu \,=\,{N\sigma\over \sqrt{2\lambda}}\,I(r_*)  \,=\, {q+\eta\over q+1} \,\frac{M}{2} \ ,
\label{nu_SUSY_c}
\eeq
which using the relation  $q=b/(2-b)$ (see (\ref{q-b})) matches (\ref{nu_SUSY_a}).

We can also use the integrated formula for $I(r)$ to write explicit expressions for $\tilde d$ and $\tilde j_y$:
\beq
\tilde d(r)\,= \tilde j_y(r) \, = \,d\,-\,\frac{(q+\eta)(2-b)}{2}\, Q \, B \left({r_*\over  r}\right)^2 \ ,
\eeq
Note that, in particular, $d=j_y$.  Notice that the non-constant terms in $\tilde d$ and  $\tilde j_y$ behave as $r^{-2}$, with no flavor corrections, which is probably a consequence of the non-anomalous dimensions of these currents.

\subsection{Nonzero temperature generalization}

Let us now consider the system at $T>0$ (\ie, when $h\not= 1$). It is possible to find a truncation of the general system of equations which defines a solution that  can be regarded as the $T>0$  generalization of the BPS system studied above. This truncation occurs when the following relations are satisfied:
\beq
E\,=\,B\,\,,
\qquad\qquad
a_0'=-h\,a_y'\ .
\label{BPS_T}
\eeq
Notice that Eq.~(\ref{constant_of_motion}) continues to be trivially satisfied.  
Moreover, Eqs.~(\ref{eom_a0_general_case}) and (\ref{eom_ax2_general_case}) still reduce to a single equation which is now:
\beq
{q+\eta\over 2b(2-q)}\partial_r
\left[
{\sqrt{q^2+b^4\,a^2}\sqrt{\Big(1-{1\over h}\Big)B^2+r^4}\over \sqrt{\Delta}}\,
\sin^2\theta\,
a_0'\right]-
B(\eta\,\cos\theta\,a'-a\,\sin\theta\,\theta')=0
\label{eom_a0_BPS_T}
\eeq
and where $\Delta$ takes the value:
\beq
\Delta\,=\,b^4r^2 h a'^{\,2}\,+\,\sin^2\theta\,\left[b^2+r^2 h\,\theta'^{\,2}\,+\,
b^2\,\Big({1\over h}-1\Big)
a_0'^{\,2}\right]\ .
\label{Delta_BPS_T}
\eeq
The equations for the flux function $a(r)$ and the embedding function $\theta(r)$ can likewise be straightforwardly  derived from the probe action. 

\section{Spectrum of mesons}
\label{mesons}

The addition of flux in the internal directions induces the breaking of parity symmetry in the Minkowski worldvolume directions. In this section we explore the effect of this parity violation on the mass spectrum of quark-antiquark  bound states which, in an abuse of language, we will refer to as mesons. The standard method to find the meson spectrum in the holographic correspondence is to analyze the normalizable fluctuations of the worldvolume gauge and scalar fields of the flavor brane.\footnote{See \cite{Hikida:2009tp,Jensen:2010vx,Zafrir:2012yg} for the calculation of the meson spectrum in the unflavored ABJM model.} Here we will restrict ourselves to analyzing the fluctuations of the gauge field $A$ around the zero-temperature supersymmetric configuration with only the internal components of $A$ are switched on. Accordingly, let us take the worldvolume gauge field  as:
\beq
A\,=\,A^{(0)}\,+\,\delta A\ ,
\eeq
where $\delta A$  is assumed to be small and the unperturbed gauge potential is given by:
\beq
A^{(0)}\,=\,L^2\,a(r)\,(d\psi+\cos\alpha\, d\beta)\ ,
\eeq
with $a(r)$ being the flux function for the SUSY embedding (\ref{a_SUSY}). We will also assume that the embedding function $\theta(r)$ does not fluctuate and is given by (\ref{theta_SUSY}).  We will denote by $f$ the first-order correction to the worldvolume field strength (\ie, $f=d\delta A$). We will assume that $f$ has only components along the $AdS$ directions. Its  components are:
\beq
f_{mn}\,=\,\partial_m \delta A_n-\partial_n\delta A_m\ ,
\eeq
where the indices $m$ and $n$ run over the $AdS$ directions. One can verify that these modes are a consistent truncation of the full set of fluctuations of the probe. In Appendix 
\ref{Fluctuations} we obtain the equations of motion for $\delta A$ by computing the first variation of the equations of motion for the probe. These equations can be written as the Euler-Lagrange equations for the following second-order  effective Lagrangian density:
\beq
{\cal L}^{(2)}=-{1\over 4}
{r^2\over (2-b)b^2}\,\left(1+(2-b)^2\,b^2\,Q^2(\cos\theta)^{2\over q}\right)
f^{mn}f_{mn}
-{Q\over 4 b\,L^4}(4-3b)
\big(\cos\theta)^{{2\over b}}
\tilde f^{mn}f_{mn}\ ,
\label{axion_QED}
\eeq
where the indices in $f^{mn}$ are raised with  the inverse of the open string metric ${\cal G}^{mn}$:
\beq
{\cal G}^{x^{\mu}\,x^{\nu}}\,=\,{\eta^{\mu\nu}\over r^2\,L^2}\,\,,
\qquad\qquad
{\cal G}^{rr}\,=\,{r^2\over L^2}\,\,{\sin^2\theta\over 1+(2-b)^2\,b^2\,Q^2\,(\cos\theta)^{2\over q}}\,\,,
\eeq
and the dual field $\tilde f^{mn}$ is defined as:
\beq
\tilde f^{\,mn}\,=\,{1\over 2}\, 
\epsilon^{mnpq}\,f_{pq}\,\,.
\label{tildef_def}
\eeq
Notice that the Lagrangian density (\ref{axion_QED}) is that of axion electrodynamics in $AdS_4$, with the axion depending on the holographic direction, showing explicitly the breaking of parity in $AdS$  when the flux is turned on.  The equation of motion derived from ${\cal L}^{(2)}$ is:
\beq
\partial_m\,\Big[r^2\Big(1+(2-b)^2\,b^2\,Q^2\,(\cos\theta)^{2\over q}\Big)\,
{\cal G}^{mp}\,{\cal G}^{nq}\,f_{pq}\,\Big]\,-\,
{\Lambda\over L^4}\,\tilde f^{rn}\,=\,0 \ ,
\label{fluct_eq_flavored}
\eeq
where $\Lambda=\Lambda(r)$ is  the function written in (\ref{calO_Lambda_explicit}). 
To solve (\ref{fluct_eq_flavored}) let us first choose the gauge in which $\delta A_r=0$, and let us separate variables in the remaining components of $\delta A$ as follows:
\beq
\delta A_{\mu}\,=\,\xi_{\mu}\,e^{ik_{\nu}\,x^{\nu}}\,R(r)\ ,
\qquad (\mu=0,1,2)\ ,
\label{ansatz_a_fluct}
\eeq
where $\xi_{\mu}$ is a constant polarization vector. The gauge condition $\delta A_r=0$, together with (\ref{fluct_eq_flavored}), imposes the following transversality condition on $\xi_{\mu}$:
\beq
k\cdot \xi\,=\,\eta^{\mu\nu}\,k_{\mu}\,\xi_{\nu}\,=\,0\ .
\label{transversality}
\eeq
When they are normalizable,  these fluctuations  are dual to vector mesons, whose mass  $m$ is given by:
\beq
m^2\,=\,-\eta^{\mu\nu}\,k_{\mu}\,k_{\nu}\ .
\label{mass_mesons}
\eeq
In order to find the equation for $R(r)$, let us choose, without loss of generality,  the Minkowski momentun $k^{\mu}$ as:
\beq
k^{\mu}\,=\,(\omega, k, 0)\ ,
\label{momentun}
\eeq
\ie, we choose our coordinates in such a way that the momentum is oriented along the $x$-direction. Notice that the mass is just $m=\sqrt{\omega^2-k^2}$. The polarization transverse to (\ref{momentun}) is:
\beq
\xi_{\mu}\,=\,\left(-{k\over \omega}\,\xi_1\,,\,\xi_1\,,\xi_2\right)\ ,
\label{polarization}
\eeq
where $\xi_1$ and $\xi_2$ are undetermined.  Let us next consider the following complex combinations of  $\xi_1$ and $\xi_2$:
\beq
\chi_{\pm}\,=\,\sqrt{1\,-\,{k^2\over \omega^2}}\,\xi_1\,\pm\,i\xi_2\ .
\label{chi_pm}
\eeq
Then, as shown in Appendix \ref{Fluctuations}, we can solve the fluctuation equation  (\ref{fluct_eq_flavored}) by taking $\chi_+\not=0$, $\chi_-=0$ or 
$\chi_+=0$, $\chi_-\not=0$ provided the radial function in the ansatz (\ref{ansatz_a_fluct}) satisfies the following ordinary differential equation:
\beq
{\cal O}\,R_{\pm}\,\pm\,m\,\Lambda\,R_{\pm}\,=\,0\ ,
\label{R_pm_eq}
\eeq
where ${\cal O}$ is the second-order differential operator defined in (\ref{calO_Lambda_explicit}) and 
$R_{+}=R_{+}(r)$ ($R_{-}=R_{-}(r)$) is the radial function for the solution with $\chi_+\not=0$ ($\chi_-\not=0$). Notice that the $\chi_+$ and $\chi_-$ modes are two helicity states which correspond to two different circular polarizations of the vector meson in the $x-y$ plane. They are exchanged by the parity transformation $\xi_2\to-\xi_2$, as  is obvious from  their definition (\ref{chi_pm}). 

\begin{figure}[ht]
\center
 \includegraphics[width=0.49\textwidth]{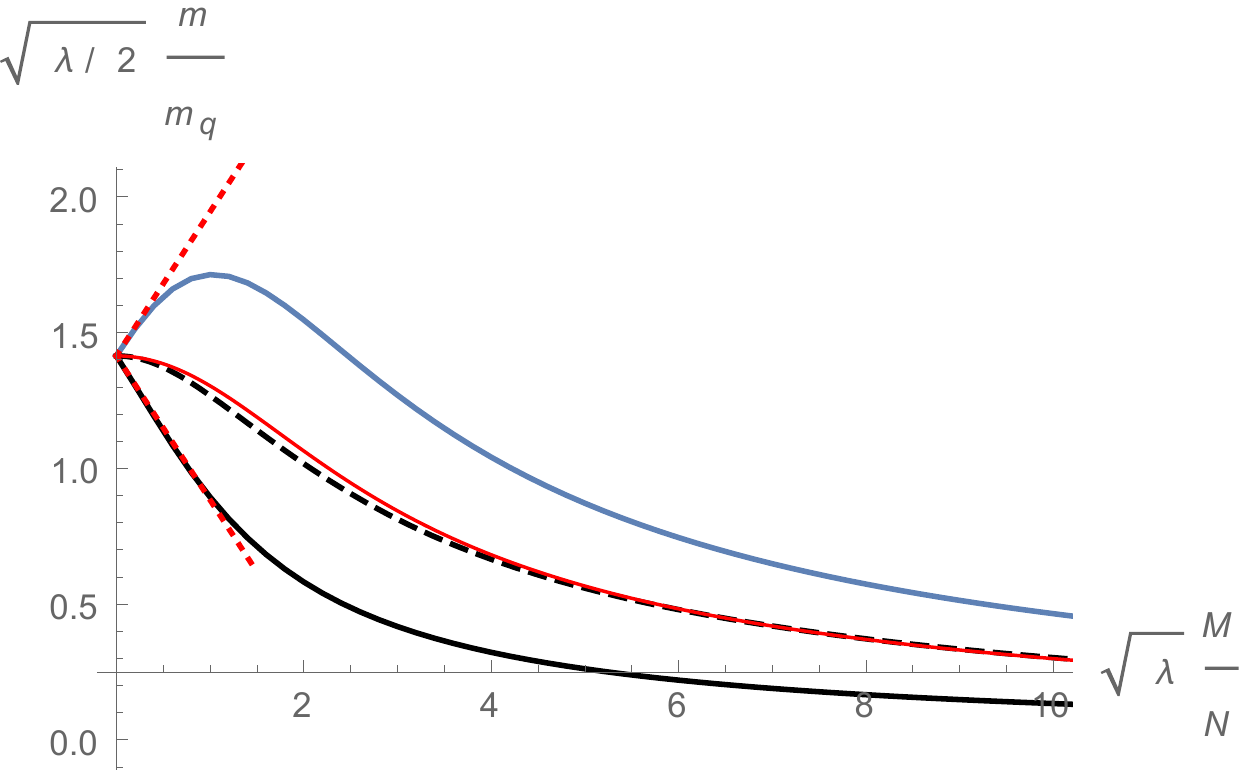}
 \qquad
 \includegraphics[width=0.45\textwidth]{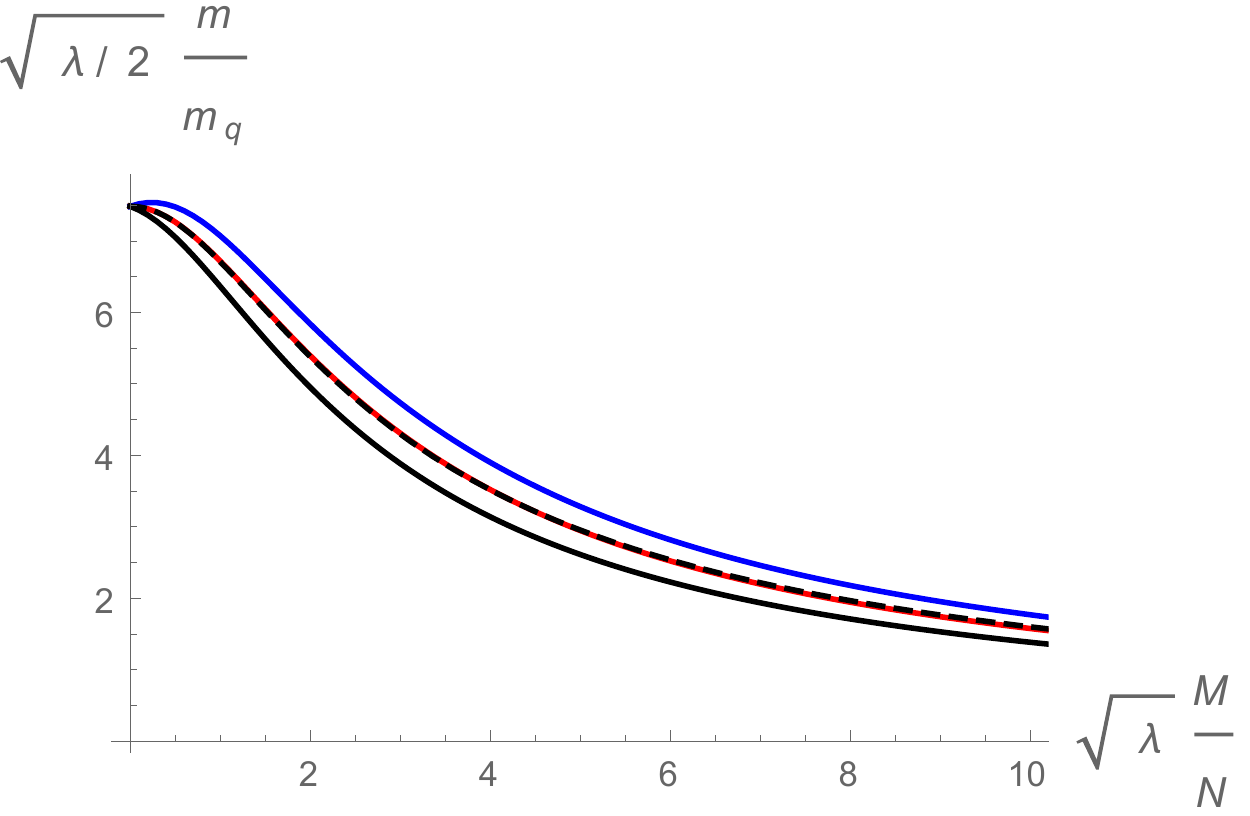}
 \caption{Meson masses  in the quenched background as a function of the flux integer $M$ for the lightest meson with excitation integer $n=0$  (left) and for $n=3$ (right). The upper blue (lower black) curves corresponds to the mode $\chi_-$ ($\chi_+$). The intermediate red curve is the average of the two curves, and the dashed black curve is the WKB estimate (\ref{WKB_averaged}). The two straight lines on the left panel are the first-order results written in  (\ref{n=0_flux_split}). }
 \label{mass_spectrum}
\end{figure}

To find the mass spectrum of the mesons we must determine the values of $m$ for which there exists a normalizable solution to (\ref{R_pm_eq}). In general this must be done numerically by using the shooting technique, although, we can make analytic estimates using the WKB approximation, which we describe in detail in Appendix \ref{Fluctuations}. The result is a tower of solutions with increasing masses which depend on the flux.  These masses depend on the location of the tip $r_*$, which can be related to the mass of the quarks $m_q$ as:
\beq
m_q\,=\,\sqrt{{\lambda\over 2}}\,\,\sigma\,r_*\ .
\eeq

In Fig.~\ref{mass_spectrum} we illustrate the dependence on the flux of the mass of the lightest state, for the quenched background with $N_f=0$. The spectra of the two helicity states are different. The mass splitting obviously vanishes at $Q=0$ and changes with the amount of  flux. 

For $N_f=0$  it is possible to compute analytically the meson mass splitting at first order in $Q$. Indeed, the fluctuation equation at first order in $Q$ for the unflavored background is:
\beq
\partial_r\,\big[\,(r^2-r_*^2)\partial_r\,R_{\pm}\,\big]\,+\,{m^2\over r^2}\,R_{\pm}\,\pm\,
2mQ\,{r_*^2\over r^3}\,R_{\pm}\,=\,0 .
\label{fluct_unflavored_orderQ}
\eeq
For $Q=0$,  Eq.~(\ref{fluct_unflavored_orderQ}) can be solved analytically in terms of hypergeometric functions. Let $R_{n}(r)$ be the normalizable regular solutions of 
(\ref{fluct_unflavored_orderQ}) for $Q=0$; they are given by:
\beq
R_{n}(r)\,=\,\Big({r_*\over r}\Big)^{2n+1}\,\,
F\left(-n-{1\over 2}, -n; 1;1-{r^2\over r_*^2}\,\right) \ ,
\qquad\qquad
n\,=\,0,1,\ldots\ .
\eeq
The corresponding  mass levels are:
\beq
{m_n\over r_*}\,=\, \sqrt{2(n+1)(2n+1)} \ ,
\qquad\qquad
n\,=\,0,1,\ldots\ .
\eeq
Let us now solve (\ref{fluct_unflavored_orderQ}) at first order in $Q$. We write:
\beq
R_{\pm,n}(r)\,=\,R_n(r)\pm Q\,\delta R_{n}(r)\,+\,{\mathcal O}(Q^2)\ ,
\eeq
where the two signs are in correspondence with the ones in (\ref{fluct_unflavored_orderQ}). The mass levels associated to $R_{\pm,n}$ will be denoted by $m_{\pm,n}$. The lightest regular normalizable modes at first order in $Q$ are given by:
\beq
R_{\pm,0}(r)\,=\,{r_*\over r}\pm Q\,
\left[{c\over r}+{1\over \sqrt{2}}\,\Big({r_*\over r}\Big)^{2}
\left(1\,-\,{r\over r_*}\,\log\left(1+{r_*\over r}\right)\right)\right]\,+\,{\mathcal O}(Q^2)\,\,,
\eeq
with $c$ being an integration constant. The masses for these modes are:
\beq
{m_{\pm,0}\over r_*}\,=\,\sqrt{2}\mp\,{3\over 4}\,Q\,+\,{\mathcal O}(Q^2)\ .
\label{n=0_flux_split}
\eeq
In Fig.~\ref{mass_spectrum} we plot these first-order results and we compare them with the numerical calculations. The first-order correction in the flux can similarly be obtained for the modes with $n\ge 1$, and the general form for the mass splitting is: 
\beq
{m_{-,n}-m_{+,n}\over r_*}\,=\,
{(2n+1)(3+4n)\over 2\,(n+1)\,\pi}\,
\left[\,{\Gamma\Big(n+{1\over 2}\Big)\over 
\Gamma\big(n+1\big)}\right]^2\,\,Q\,+\,{\mathcal O}(Q^2)\ .
\eeq
It follows from this expression that the first-order mass splitting becomes $4Q/\pi\,+\,{\mathcal O}(Q^2)$ as $n\to\infty$.

As shown in Fig.~\ref{mass_spectrum}, 
the mass averaged over the two helicities is well approximated by the WKB method:
\beq
{m_{+,n}+m_{-,n}\over 2}\approx m_{WKB}\ .
\eeq
Let us write the WKB estimate of this helicity average in the unflavored background. By
using the values of the WKB masses written  in Appendix \ref{Fluctuations}, we have:
\beq
{m_{+,n}+m_{-,n}\over 2}\approx 
{r_*\over F\Big(-{1\over 2}, {1\over 2}; 1;-Q^2\,\Big) }\,
\sqrt{2(n+1)(2n+1)}\,\,.
\label{WKB_averaged}
\eeq

\section{Discussion}
\label{discussion}

In this paper, we initiated a study of D6-brane probes with
parity-breaking flux in the ABJM background with unquenched massless
flavors. Minkowski embeddings of these probe branes holographically
described massive fermions in quantum Hall states. The filling fraction
was a half-integer in the quenched case, but received corrections when
the dynamics of the sea of massless flavors is included. The
conductivities, both in the gapped Minkowski embeddings and in the
metallic black hole ones, depended on the parity-breaking flux but also
contained a contribution from the dynamical flavors. This was interpreted
as an effect of the intrinsic disorder due to quantum fluctuations of
the fundamental degrees of freedom.

Despite the complexity of the equations of motion we managed to obtain
an explicit, analytic family of supersymmetric solutions with nonzero
charge density, electric, and magnetic fields. For these gapped QH
solutions, we obtained an analytic expression for the Hall
conductivity, which includes the effects of quark loops. We also
analyzed the residual $SO_+(1,1)$ boost invariance of the system at zero
temperature; this is a powerful tool for generating non-supersymmetric
solutions with general electric and magnetic fields starting from
solutions with either $E$=0 or $B$=0. We also explored the effect of
the parity violation on the computation of the meson spectrum. We
restricted our analysis to the fluctuations of the gauge field around
the zero-temperature supersymmetric configuration in which only the internal gauge
field components were switched on.

There remain many open topics for further investigation.  Although, we presented here a set of analytic, BPS solutions for a very specific set of parameters, more general solutions with nonzero temperature and arbitrary $d$, $Q$, $E$, and $B$ still need to be studied, probably numerically.  A thorough analysis of the thermodynamics and phase structure is needed to provide a complete understanding of this model. For example, we anticipate a phase transition as the temperature is increased from the MN to the black hole embeddings.\footnote{For a massless probe brane embedding, this phase transition is related to the breaking of chiral symmetry.  See {\it e.g.}~\cite{Preis:2010cq, Preis:2012fh, Filev:2007gb, Filev:2010pm} for the holographic realization of (inverse) magnetic catalysis and \cite{Filev:2011mt, Erdmenger:2011bw, Jokela:2013qya} for the study of the effect of the dynamical flavors on magnetic catalysis.}  

The effects of internal flux and the sea of massless quarks are particularly interesting.  And, we would also like to understand how the BPS solutions fit in to the complete picture. 

Because there are so many possible parameters to vary, it makes sense to start by isolating one or two.  A good first step could be to analyze, along the lines of \cite{Jokela:2012dw}, the thermodynamics of the D6-brane probe with only the internal flux, presented in Section \ref{Full_ansatz}.  In the absence of massless flavors, this system is essentially a probe D6-brane in the ABJ background, but with zero worldvolume gauge field.  This is another system which deserves a detailed thermodynamic study.

Another interesting problem for the future is the study of flavor branes with internal flux in the ABJM background with unquenched massive quarks presented in \cite{Bea:2013jxa}. The geometry found in \cite{Bea:2013jxa} is a running solution flowing between two $AdS$ spaces,  in  which the control parameter is the mass of the sea quarks. One expects to find conductivities depending on the mass of the dynamical quarks, which interpolate between the values found here (for massless sea quarks) and the unflavored values (for infinitely massive quarks).

In this paper we considered brane probes with electric and magnetic fields in their worldvolume and we have neglected the backreaction of these electromagnetic fields. Computing this backreaction is a very complicated task. One possible intermediate step could be considering the geometry dual to the non-commutative ABJM model, which was found in \cite{Imeroni:2008cr} by applying a TsT duality transformation. By  adding internal flux to the probe branes we should be able to find Hall states similar to those found here \cite{work_in_progress}.

We initiated a very limited fluctuation analysis in Section \ref{mesons}, and a more thorough study is needed.  One goal would be to compute the full meson dispersion about the BPS solution of Section \ref{BPS-sol}.  The lightest neutral excitations of QH fluids are magneto-rotons, collective excitations whose minimum energy is at nonzero momentum.  They have been detected in experiments, for example \cite{rotonexp}, and they have also been found in other holographic probe-brane QH models \cite{Jokela:2010nu, Jokela:2011sw}.  Naturally, we would like to know if the spectrum of the model presented here also includes rotons.

Homogeneity-breaking instabilities seem to be a general feature of black hole embeddings in related brane models \cite{BallonBayona:2012wx, Bergman:2011rf, Jokela:2012se, Jokela:2012vn}, which are examples of the general type of instability described in \cite{Nakamura:2009tf}.  In some cases, examples of spatially-modulated ground states have been found explicitly; for example, see \cite{Rozali:2012es, Withers:2013loa, Withers:2014sja, Jokela:2014dba}.  A thorough analysis of the quasi-normal mode spectrum is needed to determine whether such instabilities exists in this model.  If so, the ABJM system, due to its symmetries and other special properties, might afford an ideal laboratory to study inhomogeneous phases.

Another interesting area to explore is alternative quantization of the D6-brane worldvolume gauge field.  In a four-dimensional bulk, the gauge field can take Dirichlet, Neumann, or mixed boundary conditions in the UV, and these choices correspond to different boundary CFTs dual to strongly coupled anyon fluids in dynamical gauge fields. In particular, by changing the quantization as in \cite{Jokela:2013hta, Brattan:2013wya, Jokela:2014wsa}, this ABJM system could be turned from a quantum Hall fluid into an anyon superfluid.

\vspace{0.2cm}

{\bf \large Acknowledgments}

Y. ~B and A.~V.~R. are funded by the Spanish grant FPA2011-22594, by the Consolider-Ingenio 2010 Programme CPAN (CSD2007-00042), by Xunta de
Galicia (Conselleria de Educaci\'on, grant INCITE09-206-121-PR and grant PGIDIT10PXIB206075PR), and by FEDER.
Y.~B. is also supported by the foundation Pedro Barrié de la Maza.
N.~J. is supported by the Academy of Finland grant no. 1268023. 
D.~Z. is funded by the FCT fellowship SFRH/BPD/62888/2009. 
M.~L. is supported by funding from the European Research Council under the European Union's Seventh Framework Programme (FP7/2007-2013) / ERC Grant agreement no.~268088-EMERGRAV. This work is part of the $\Delta$-ITP consortium and also supported in part by the Foundation for Fundamental Research on Matter (FOM), both are parts of the Netherlands Organization for Scientific Research (NWO) that is funded by the Dutch Ministry of Education, Culture and Science (OCW).  
 Y.~B. wants to thank the University of Swansea for hospitality at the final stages of this work.
N.~J. wishes to thank the hospitality of the Universidade de Santiago de Compostela, especially for the chipirones, while this work was in progress.  
M.~L. would like to thank the University of Helsinki and the Helsinki Institute of Physics for the very warm hospitality, and especially for the reindeer, while this work was being completed.

\newpage

\appendix
\vskip 1cm
\renewcommand{\theequation}{\rm{A}.\arabic{equation}}
\setcounter{equation}{0}

\section{Details of the background geometry}
\label{Background_details}

In this appendix we specify the coordinate system we employ to represent the metric and forms of the background. Let us begin with the four-sphere part of the internal metric (\ref{internal-metric-flavored}). Let $\omega^i$ ($i=1,2,3$) be the $SU(2)$ left-invariant one-forms which satisfy $d\omega^i={1\over2}\,\epsilon_{ijk}\,\omega^j\wedge\omega^k$. Together with a new coordinate $\alpha$, the $\omega^i$'s can be used to parameterize the metric of  the four-sphere ${\mathbb S}^4$ as:
\beq
ds^2_{{\mathbb S}^4}\,=\,d\alpha^2\,+\,
{\sin^2\alpha\over 4}
\left[(\omega^1)^2+(\omega^2)^2+(\omega^3)^2
\right]\ ,
\label{S4metric}
\eeq
where $0\le \alpha<\pi$. The $SU(2)$ instanton one-forms $A^i$  which fiber the ${\mathbb S}^2$ over the $ {\mathbb S}^4$ in (\ref{internal-metric-flavored}) can be written in these coordinates as:
\beq
A^{i}\,=\,-\sin^2\left({\alpha\over 2}\right)
\,\,\omega^i\,\,. 
\label{A-instanton}
\eeq
Let us next parametrize the $z^i$ coordinates of the ${\mathbb S}^2$ by means of two angles $\theta$ and $\varphi$ ($0\le\theta<\pi$, $0\le\varphi<2\pi$), namely:
\beq
z^1\,=\,\sin\theta\,\cos\varphi\,\,,\qquad\qquad
z^2\,=\,\sin\theta\,\sin\varphi\,\,,\qquad\qquad
z^3\,=\,\cos\theta\,\,.
\label{cartesian-S2}
\eeq
Then, one can easily prove that the  $ {\mathbb S}^2$ part of the metric 
(\ref{internal-metric-flavored}) can be written as:
\beq
\left(d x^i\,+\, \epsilon^{ijk}\,A^j\,z^k\,\right)^2\,=\,(E^1)^2\,+\,(E^2)^2\ ,
\eeq
where  $E^1$ and $E^2$ are the following one-forms:
\bear
E^1&=&d\theta+\sin^2\big({\alpha\over 2}\big)\,
\left[\sin\varphi\,\omega^1-\cos\varphi\,\omega^2\right]
\rc\rc
E^2&=&\sin\theta\left[d\varphi-\sin^2\big({\alpha\over 2}\big)
\,\omega^3\right]+\sin^2\big({\alpha\over 2}\big)\,
\cos\theta\left[\cos\varphi\,\omega^1+\sin\varphi\,\omega^2\right]\ .
\label{Es}
\eear

In order to write the explicit expression for $F_2$,  we  first define three new one-forms  $S^i$ $(i=1,2,3)$ as the following rotated version of the $\omega^i$'s:
\bear
S^1&=&\sin\varphi\,\omega^1-\cos\varphi\,\omega^2 \rc\rc
S^2&=&\sin\theta\,\omega^3-\cos\theta\left(\cos\varphi\,\omega^1+
\sin\varphi\,\omega^2\right) \rc\rc
S^3&=&-\cos\theta\,\omega^3-\sin\theta\left(\cos\varphi\,\omega^1+
\sin\varphi\,\omega^2\right)\ .
\label{rotomega}
\eear 
Next, we define the one-forms ${\cal S}^{\alpha}$   and ${\cal S}^{i}$ as:
\beq
{\cal S}^{\alpha}\,=\,d\alpha\,\,,\qquad\qquad
{\cal S}^{i}\,=\,{\sin\alpha\over 2}\,S^i \,\,,\qquad(i=1,2,3)\ ,
\label{calS}
\eeq
in terms of which the metric of the four-sphere is just 
$ds^2_{{\mathbb S}^4}=({\cal S}^{\alpha})^2+\sum_i({\cal S}^{i})^2$.  

With these definitions,  the ansatz for the RR two-form $F_2$ for the flavored background  is
\beq
F_2\,=\,{k\over 2}\,\,\Big[\,\,
E^1\wedge E^2\,-\,\eta\,
\big({\cal S}^{\alpha}\wedge {\cal S}^{3}\,+\,{\cal S}^1\wedge {\cal S}^{2}\big)
\,\,\Big]\ .
\label{F2-flavored}
\eeq
Note that the two-form $F_2$ is not closed when $\eta\not=1$; $dF_2$ is proportional to the charge distribution three-form of the flavor D6-branes. 
The RR four-form $F_4$ is:
\beq
F_4\,=\,{3k\over 4}\,\,\,{(\eta+q)b\over 2-q}\,\,L^2\,\,\Omega_{BH_4}\ ,
\eeq
where $\Omega_{BH_4}$ is the volume-form of the four-dimensional black hole (\ref{BH4-metric}). The solution is completed by a constant dilaton $\phi$ given by
\beq
e^{-\phi}\,=\,{b\over 4}\,{\eta+q\over 2-q}\,{k\over L}\ . 
\label{dilaton-flavored-squashings}
\eeq

Let us now spell out the embedding of the D6-brane probe in the background geometry. We first represent the $SU(2)$ left-invariant one-forms $\omega^i$ in terms of three angles $\hat\theta$, $\hat\varphi$, and $\hat\psi$ as:
\bear
\omega^1 & = & \cos\hat\psi\,d\hat\theta+\sin\hat\psi\,\sin\hat\theta\,d\hat\varphi \rc
\omega^2 & = & \sin\hat\psi\,d\hat\theta-\cos\hat\psi\,\sin\hat\theta\,d\hat\varphi \rc
\omega^3 & = & d\hat\psi+\cos\hat\theta \,d\hat\varphi\,\,.
\label{w123}
\eear
In these coordinates our embedding is defined by the conditions:
\beq
\hat\theta\, ,\ \hat\varphi \,=\,{\rm constant}\ ,
\label{RP3-cycle}
\eeq
with the coordinate $\theta$ defined in (\ref{cartesian-S2}) being a function of the radial coordinate $r$.  The relation of the coordinates defined here with those used in (\ref{induced_metric}) to write the internal part of the induced metric is as follows: The angle $\alpha$ here is equal to the one introduced in (\ref{S4metric}), while $\beta$ and $\psi$ are given by
\beq
\beta\,=\,{\hat\psi\over 2}\,\,,\qquad\qquad
\psi\,=\,\varphi\,-\,{\hat\psi\over 2}\ .
\label{RP3-angles}
\eeq
It is now easy to check that the pullback of the metric (\ref{flavoredBH-metric}) to the worldvolume is, indeed, the line element written in (\ref{induced_metric}).

\subsection{Matching the unflavored ABJ model}
\label{matching_ABJ}

Let us now explore our prescription (\ref{wv_quantization}) in the case of the unflavored model. The main point is that, when $N_f=0$, the worldvolume gauge field for the supersymmetric configurations can be understood as induced by a flat NSNS $B_2$ field of the bulk, which is proportional to the K\"ahler form $J$ of ${\mathbb C}{\mathbb P}^3$. When the coefficient multiplying 
$J$ is appropriately quantized, the corresponding supergravity solution is the dual of the ABJ theory with gauge group $U(N+M)_{k}\times U(N)_{-k}$. We will see that the rank difference $M$ can be identified with the quantization integer in (\ref{wv_quantization}).  

Let us begin our analysis by writing the K\"ahler form of  ${\mathbb C}{\mathbb P}^3$ in our variables:
\beq
J\,=\,{1\over 4}\,\,\Big(\,E^1\wedge E^2\,-\,\big(
{\cal S}^{\alpha}\wedge {\cal S}^{3}\,+\,{\cal S}^1\wedge {\cal S}^{2}\big)\,\Big)\ .
\label{Kahler}
\eeq
The pullback of $J$  to the probe brane worldvolume is:
\beq
\hat J\,=\,{\theta'\,\sin\theta\over 4}\,dr\wedge \big[\,
d\psi+\cos\alpha\,d\beta\,\big]\,+\,{\cos\theta\sin\alpha\over 4}\,d\alpha\wedge d\beta\ ,
\eeq
and, as claimed, it has the form written in  (\ref{F_internal}) if we identify the flux function $a(r)$ with:
\beq
a(r)\,=-Q\,\cos\theta(r)\ ,
\label{a_theta_unflavored}
\eeq
where $Q$ is a constant. Actually, one can check that the worldvolume gauge field $F$ (\ref{F_internal}) for this unflavored case can be written in terms of the pullback of $J$ as:
\beq
F\,=\,4\,L^2\,Q\,\hat J\,\,.
\label{internal_flux}
\eeq
We will see below  in Appendix \ref{kappa} that the relation (\ref{a_theta_unflavored}) between the flux and embedding functions is dictated by supersymmetry when $N_f=0$. Notice also that the flux function at the tip is just $Q$, as in (\ref{a_*_Q}).
 
In the DBI+WZ action for D-branes, the worldvolume gauge field $F$ is always combined additively with the pullback of $B_2$. It follows that,  in this case, the worldvolume flux can alternatively be thought of as induced by the following NSNS $B_2$ field:
\beq
B_2\,=\,4\,Q\,L^2\,J\,\,.
\label{B2-para}
\eeq
Notice that $B_2$ is a closed two-form, and it has the same form as in the proposed gravity dual of the ABJ model \cite{Aharony:2008gk} with gauge group $U(N+M)_{k}\times U(N)_{-k}$, where $Q$ is related to $M$. Actually, the integer $M$ is determined by the discrete holonomy of $B_2$ on the ${\mathbb C}{\mathbb P}^1$ cycle of the ${\mathbb C}{\mathbb P}^3$ space, which is inherited from the holonomy of the three-dimensional three-form potential of the eleven dimensional supergravity along the torsion cycle of the ${\mathbb S}^7/{\mathbb Z_k}$. Let us compute explicitly the integral of the 
two-form (\ref{B2-para}) along the ${\mathbb C}{\mathbb P}^1$.  In our coordinates (see \cite{Conde:2011sw}) the  ${\mathbb C}{\mathbb P}^1$ is obtained by keeping the coordinates of the ${\mathbb S}^4$ cycle fixed. Therefore, the pullback of $J$ is just:
\beq
J_{ |_{{\mathbb C}{\mathbb P}^1}}\,=\,{1\over 4}\,\sin\theta\,d\theta\wedge d\varphi\,\,,
\eeq
and thus the integral of $J$ along the ${\mathbb C}{\mathbb P}^1$ is:
\beq
\int_{{\mathbb C}{\mathbb P}^1}\,J\,=\,\pi\,\,.
\eeq
It follows from (\ref{B2-para})  that:
\beq
\int_{{\mathbb C}{\mathbb P}^1}\,B_2\,=\,4\pi\,L^2\,Q\,\,.
\eeq
Let us now use our quantization condition (\ref{a*_quantization}) and the identification 
(\ref{a_*_Q}) to write the period of $B_2$ in terms of $k$ and  the quantization integer $M$. We get:
\beq
\int_{{\mathbb C}{\mathbb P}^1}\,B_2\,=\,(2\pi)^2\,{M\over k}\,\,,
\eeq
which is the fractional holonomy proposed in \cite{Aharony:2008gk} for the gravity dual of the $U(N+M)_{k}\times U(N)_{-k}$ theory. 

The coefficient $Q$ can also be fixed by looking at the Page charge  $Q_4$ for fractional D2-branes (D4-branes wrapped on a ${\mathbb C}{\mathbb P}^1$ two-cycle), which is given by the following integral over the ${\mathbb C}{\mathbb P}^2$ dual to the   ${\mathbb C}{\mathbb P}^1$  where the D4-branes are wrapped:
\beq
Q_4\,=\,{1\over (2\pi)^3}\,\int_{{\mathbb C}{\mathbb P}^2}\,\Big[F_4+B_2\wedge F_2]
\,\,. 
\label{Q_4}
\eeq
We require that $Q_4$ is equal to our quantization integer $M$, which can then be interpreted as the number of fractional D2-branes. Taking into account (\ref{B2-para}) and that  $F_2=2k\,J$ for this unflavored case, we get:
\beq
Q_4\,=\,{k\,L^2\,Q\over  \pi^3}\,\int_{{\mathbb C}{\mathbb P}^2}\,
J\wedge J\,\,.
\eeq
To compute this integral we use the fact that the equation of the
${\mathbb C}{\mathbb P}^2$ cycle in our coordinates is $\varphi=\theta=\pi/2$ (see Appendix A in \cite{Conde:2011sw}), which implies:
\beq
J\wedge J_{ |_{{\mathbb C}{\mathbb P}^2}}\,=\,{1\over 16}\,
\sin^2{\alpha\over 2}\,\sin\alpha\,
\,d\alpha\wedge \omega^{1}\wedge \omega^{2}\,\wedge \omega^{3}\,\,.
\eeq
Then, it follows that:
\beq
\int_{{\mathbb C}{\mathbb P}^2}\,J\wedge J\,=\,\pi^2\,\,.
\eeq
Therefore,
\beq
Q_4\,=\,{k\,L^2\,Q\over  \pi}\,\,,
\eeq
and the quantization condition $Q_4=M$ coincides with the one obtained in (\ref{a*_quantization}) for $a_*\,=\,-Q$. 

\renewcommand{\theequation}{\rm{B}.\arabic{equation}}
\setcounter{equation}{0}

\section{Probe brane equation of motion}
\label{EOMs}

Let us consider a D$p$-brane probe propagating in a background of type II supergravity.  Let $g_{ij}$ denote the components of the induced metric on the worldvolume:
\beq
g_{ij}\,=\,g_{mn}\,\partial_i\,X^{m}\,\partial_j\,X^n\,\,,
\eeq
where the $X^{n}$ are coordinates of the ten-dimensional space and  $g_{mn}$ is the target space metric of the background. 
In what follows $m,n,\ldots$ will denote indices of the target space, whereas $i,j,\ldots$ will represent worldvolume indices. Let us denote by $M$ the following matrix:
\beq
M\,=\,g+F\,\,,
\eeq
where $F=dA$ is the worldvolume gauge field. Then, the action of a D$p$-brane probe can be written as:
\beq
S_{D_p}\,=\,S_{DBI}+S_{WZ}\,\,,
\eeq
where the DBI and WZ terms are:
\beq
S_{DBI}\,=\,-T_{Dp}\,\int_{{\cal M}_{p+1}}\,d^{p+1}\xi\,\,e^{-\phi}\,
\sqrt{-\det M}\,\,,
\qquad\qquad
S_{WZ}\,=\,T_{Dp}\,\int_{{\cal M}_{p+1}}\,e^{F}\wedge C\,\,,
\eeq
with $T_{Dp}$ being the D$p$-brane tension (from now on  in this appendix we will take $T_{Dp}=1$) and $C=\sum_r C_r$ is the sum of RR potentials. 
In order to write the equations of motion derived from this action following the analysis of Section 2 of \cite{Skenderis:2002vf}, let us consider the inverse $M^{-1}=[M^{ij}]$  of the matrix $M=[M_{ij}]$ and let us decompose $M^{-1}$ in its symmetric and antisymmetric parts as:
\beq
M^{-1}\,=\,{\cal G}^{-1}\,+\,{\cal J}\,\,,
\eeq
where ${\cal J}=[{\cal J}^{ij}]$ is the antisymmetric component of $M^{-1}$ and ${\cal G}^{-1}=[{\cal G}^{ij}]$ is the inverse open string metric. Then,  the equation of motion of the gauge field component $A_j$ is \cite{Skenderis:2002vf}:
\beq
\partial_j\Big(e^{-\phi}\,\sqrt{-\det M}\,{\cal J}^{ji}\Big)\,=\,j^{i}\,\,,
\label{eom_gauge_general}
\eeq
where the source current for the gauge field $j^{i}$ is given by:
\beq
j^{i}\equiv {\delta S_{WZ}\over \delta A_i}\,\,.
\eeq
Moreover, the equation for the scalar field $X^m$ becomes \cite{Skenderis:2002vf}:
\bear
&&-\partial_i\Big(e^{-\phi}\,\sqrt{-\det M}\,{\cal G}^{ij}\,\partial_j\,
X^n\,g_{nm}\Big)\, \,\rc\rc
&&\qquad\qquad
+\sqrt{-\det M}\,\Big({e^{-\phi}\over 2}\,
{\cal G}^{ji}\,\partial_i\,X^n\,\partial_j\,X^p\,g_{np,m}\,-\,e^{-\phi}\,\partial_m\,\phi\Big)\,=\,
j_m\,\,,\qquad\qquad
\label{eom_scalars_general}
\eear
where the source for the scalar $X^m$ is:
\beq
j_m\equiv {\delta S_{WZ}\over \delta X^m}\,\,.
\eeq

\subsection{Currents for the D6-brane}

Let us  write the form of the currents for the case of a D6-brane probe. In this case,
the WZ term of the action is:
\beq
S_{WZ}\,=\,\int_{{\cal M}_7}\,\left(\hat C_7\,+\,\hat C_5\wedge F\,+\,{1\over 2}\,
\hat C_3\wedge F\wedge F\,+\,{1\over 6}\hat C_1\wedge F\wedge F\wedge F
\right)\ .
\label{WZ-D6}
\eeq
Let us perform a general variation of the worldvolume gauge field $F\to F+d(\delta A)$, under which $S_{WZ}$ varies as:
\beq
\delta S_{WZ}\,=\,\int_{{\cal M}_7}\,
\left( \hat C_5\,+\,\hat C_3\wedge F\,+\,{1\over 2}\,\hat C_1\wedge F\wedge F\right)
\wedge d(\delta A)\ .
\eeq
In order to compute the current associated to the worldvolume gauge field, we use the fact that,  for any odd-dimensional form ${\cal O}$, one has
\beq
{\cal O}\wedge d(\delta A)\,=\,d{\cal O}\wedge \delta A\,-\,d({\cal O}\wedge \delta A)\ .
\eeq
The total derivative generates a boundary term which vanishes since we are assuming that $A$ is fixed at the boundary in the variational process.\footnote{Note that, although we have chosen Dirichlet boundary conditions for $A$ here, $A$ can, in fact, have arbitrary mixed boundary conditions, corresponding to alternative quantization, as discussed in \cite{Jokela:2013hta}.} Taking into account that, with our notation $F_4=-dC_3$, we get:
\beq
\delta S_{WZ}\,=\,\int_{{\cal M}_7}\,
\left( \hat F_6\,-\,\hat F_4\wedge F\,+\,{1\over 2}\,\hat F_2\wedge F\wedge F\right)
\wedge \delta A\ .
\eeq
Then, the gauge current along the worldvolume direction $i$ is given by the expression:
\beq
j^i\,d^7\xi\,=\,\left(\hat F_6\,-\,\hat F_4\wedge F\,+\,{1\over 2}\,\hat F_2\wedge F\wedge F\right)\wedge d\xi^i\ .
\eeq

In order to compute the source current $j_m$ for $X^m$, we should vary in (\ref{WZ-D6}) the  scalars which enter the pullback of the RR potentials. It turns out that the final expression can be written in a rather simple form, which we will now spell out. 
Let $V=V^{m}\,{\partial \over \partial X^m}$ be any vector field in target space. The interior product of $V$ with a $p$-form $\omega$ is a $(p-1)$-form $ \iota_V\omega$ defined as follows. Let $\omega$ be:
\beq
\omega\,=\,{1\over p!}\,\omega_{n_1,\ldots ,n_p}\,
dX^{n_1}\wedge\cdots\wedge dX^{n_p}\,\,.
\eeq
Then, $ \iota_V\omega$  is given by:
\beq
\iota_V\omega\,=\,{1\over (p-1)!}\,V^{m}\,
\omega_{m,m_1,\ldots, m_{p-1}}\,dX^{m_1}\wedge \cdots\wedge dX^{m_{p-1}}\,\,.
\eeq
Let $\iota_m\omega$ denote the interior product of $\omega$ and the vector $\partial/\partial X^m$:
\beq
\iota_m\omega\,\equiv\,\iota_{{\partial\over \partial X^m}}\omega\,\,. 
\eeq
Then, the current $j_m$, corresponding to the scalar $X^{m}$,  can be written as:
\beq
j_m\,d^7\xi\,=\,\widehat{\iota_m F_8}\,+\,\widehat{\iota_m F_6}\wedge F\,-\,{1\over 2}\,
\widehat{\iota_m F_4}\wedge F\wedge F\,+\,{1\over 6}\,
\widehat{\iota_m F_2}\wedge F\wedge F\wedge F\,\,,
\label{scalar_current}
\eeq
where the hat denotes the pullback of the different $\iota_m F_r$ to the worldvolume. In (\ref{scalar_current}) $F_8$ and $F_6$ are defined as  Hodge duals of $F_2$ and $F_4$, respectively, \ie, $F_8=-*F_2$ and $F_6=-*F_4$.

Notice that we have derived the expressions of $j^i$ and $j_m$ from the action (\ref{WZ-D6}), where we have assumed the existence of the RR potentials $C_r$.  In the case of backreacting flavor some Bianchi identities are  violated and, as a consequence, some of the RR potentials do not exist. However, the currents  $j^i$ and $j_m$ (and the corresponding equations of motion) only depend on the RR field strengths and their pullbacks, and then they can be generalized to the case in which we include the backreaction. This is the point of view we will adopt in what follows.

\subsection{The equations of motion for our ansatz}

We now write explicitly the equations of motion for the D6-brane with a gauge potential $A$ as the one written in (\ref{A_full_ansatz}). We will also assume that the embedding is defined by the conditions (\ref{RP3-cycle}) with $\theta=\theta(r)$ being a function of $r$ to be determined. The set of worldvolume coordinates we will employ is:
\beq
\xi^{i}\,=\,(t, x, y , r, \alpha,\beta,\psi)\,\,,
\eeq
where $\alpha$, $\beta$, and $\psi$ are the angles defined in (\ref{S4metric}) and (\ref{RP3-angles}).  First of all, let us write the non-zero components of the worldvolume gauge field strength $F$ corresponding to the potential  (\ref{A_full_ansatz}):
\bear
&&F_{t\,x}\,=\,L^2\,E\,\,,\qquad\qquad
F_{x\,y}\,=\,L^2\,B\,\,,\rc\rc
&&F_{r\,t}\,=\,L^2\,a_0'\,\,,\qquad\qquad
F_{r\,x}\,=\,L^2\,a_x'\,\,,\qquad\qquad
F_{r\,y}\,=\,L^2\,a_y'\,\,,\rc\rc
&&F_{r\,\psi}\,=\,L^2\,a'\,\,,\qquad\qquad
F_{r\,\beta}\,=\,L^2\,a'\cos\alpha\,\,,\qquad\qquad
F_{\alpha\,\beta}\,=\,-L^2\,a\,\sin\alpha\,\,,
\qquad\qquad\qquad
\eear
where the prime denotes the derivative with respect to the radial variable. 
Notice that, in our ansatz, isotropy in the $x-y$ plane is explicitly broken by the electric field in the $x$-direction.

We will start by computing the different components of the currents appearing in (\ref{eom_gauge_general}) and (\ref{eom_scalars_general}). It is straightforward  to prove that $\hat F_6=0$, and it therefore does not contribute to 
$j^i$ and $j_m$.  The non-vanishing components of the gauge current $j^i$ are:
\bear
&&j^t\,=\,{kL^4\over 2}\,B\,\sin\alpha\,(\eta\,\cos\theta\,a'\,-\,a\,\sin\theta\,\theta')\,\,,\rc\rc
&&j^{y}\,=\,{kL^4\over 2}\,E\,\sin\alpha\,(\eta\,\cos\theta\,a'\,-\,a\,\sin\theta\,\theta')\,\,,\rc\rc
&&j^{\psi}\,=\,{kL^4\over 2}\,\sin\alpha\,\Bigg({3b\over 2}\,{\eta+q\over 2-q}\,r^2\,a\,-\,
\eta\,(B\,a_0'+E\,a_y')\cos\theta \Bigg)\,\,.
\label{gauge_current_comps_general}
\eear
We now work out the current for the three transverse scalars.  First we compute the interior products of $F_8$ with the tangent vectors along the three scalar directions $m=\theta, \hat\theta, \hat\varphi$. We find:
\beq
\widehat{\iota_{\theta} F_8}\,=\,-{(3-2b)(q+\eta)q\over 8 b^3 (2-q)}\,kL^6\,\sin
\alpha\,r^2\,\sin(2\theta)\,\,d^7\xi\,\,.
\label{iF8_scalars}
\eeq
Moreover,  the product of $F_8$  with the other two tangent vectors gives a result proportional  to $\widehat{\iota_{\theta} F_8}$:
\beq
\widehat{\iota_{\hat\theta} F_8}\,=\,-\sin^2\Big({\alpha\over 2}\Big)\,\sin(\beta-\psi)\,
\widehat{\iota_{\theta} F_8}\,\,,
\qquad
\widehat{\iota_{\hat\varphi} F_8}\,=\,\sin\hat\theta\sin^2\Big({\alpha\over 2}\Big)\,\cos(\beta-\psi)\,
\widehat{\iota_{\theta} F_8}\,\,.
\label{iF8_hat_scalars}
\eeq
We already mentioned that $F_6$ does not contribute since its pullback is zero. It is also immediate to check that $F_4$  does not have components along the transverse scalars and will not contribute to $j_m$. The contribution of $F_2$ to $j_{\theta}$ is determined by:
\beq
{1\over 6}\,
\widehat{\iota_{\theta} F_2}\wedge F\wedge F\wedge F\,=\,{kL^6\over 2}\,
\sin\alpha\,a\,\sin\theta\,(B\,a_0'\,+\,E\,a_y')
\,\,d^7\xi\,\,,
\eeq
while the result for the other scalars are:
\bear
&&\widehat{\iota_{\hat\theta} F_2}\wedge F\wedge F\wedge F\,=\,
-\sin^2\Big({\alpha\over 2}\Big)\,\sin(\beta-\psi)\,
\widehat{\iota_{\theta} F_2}\wedge F\wedge F\wedge F\,\,,\rc\rc
&&\widehat{\iota_{\hat\varphi} F_2}\wedge F\wedge F\wedge F\,=\,
\sin\hat\theta\sin^2\Big({\alpha\over 2}\Big)\,\cos(\beta-\psi)\,
\widehat{\iota_{\theta} F_2}\wedge F\wedge F\wedge F\,\,.
\label{iF2_hat_scalars}
\eear
Notice that the proportionality factors in (\ref{iF8_hat_scalars}) and (\ref{iF2_hat_scalars}) are the same. Thus, the current for the scalar $\theta$ becomes:
\beq
j_{\theta}\,=\,-{kL^6\over 2}\,\sin\alpha\,\sin\theta\,
\Bigg({(3-2b)(q+\eta)q\over 2\, b^3 (2-q)}\,r^2\cos\theta\,-\,a\,(B\,a_0'\,+\,E\,a_y')\,\Bigg)\,\,.
\label{j_theta_general_case}
\eeq
Moreover, the other two components of $j_m$ are:
\beq
j_{\hat\theta}\,=\,-\sin^2\Big({\alpha\over 2}\Big)\,\sin(\beta-\psi)\,j_{\theta}\,\,,
\qquad
j_{\hat\varphi}\,=\,\sin\hat\theta\,\sin^2\Big({\alpha\over 2}\Big)\,\cos(\beta-\psi)\,
\,j_{\theta}\,\,.
\label{j_hat_scalars_exp}
\eeq

Let us now  finally write the equations of motion  for the different gauge field components and scalars. We have to compute the different components of the antisymmetric tensor ${\cal J}^{ij}$, as well as the elements of the inverse open string metric ${\cal G}^{ij}$.  This calculation is straightforward (although rather tedious in some cases) and we limit ourselves to give the final result for the equations. The equation of $A_t$ is:
\bear
&&{q+\eta\over 2b(2-q)}\,\partial_r
\Bigg[{\sqrt{h}\,\sqrt{q^2+b^4\,a^2}\over \sqrt{\Delta}
\sqrt{(B^2+r^4)h-E^2}}\,\sin^2\theta\,
\big[B(B\,a_0'\,+\,E\,a_y')\,+r^4\,a_0'\big]
\Bigg]
\,\,\rc\rc
&&\qquad\qquad\qquad\qquad
-B(\eta\,\cos\theta\,a'-a\,\sin\theta\,\theta')\,=\,0\,\,,
\label{eom_a0_general_case}
\eear
where $\Delta$ is the quantity defined in (\ref{Delta_def_general}).  The equation for 
 $A_{x}$ can be integrated once ($a_x$ is a cyclic variable) to give the following equation for $a_x'$:
\beq
r^4\,h^{{3\over 2}}\,\sin^2\theta\,
{\sqrt{q^2+b^4\,a^2}\,a_x'\over 
\sqrt{\Delta}\,\sqrt{(B^2+r^4)h-E^2}}\,=\,
{\rm constant}\,\,.
\label{eom_ax_general_case}
\eeq
The equation for $A_{y}$ is also non-trivial and given by:
\bear
&&{q+\eta\over 2b(2-q)}\,\partial_r
\Bigg[{\sqrt{h}\,\sqrt{q^2+b^4\,a^2}\over \sqrt{\Delta}
\sqrt{(B^2+r^4)h-E^2}}\,\sin^2\theta\,
\big[E(B\,a_0'\,+\,E\,a_y')\,-r^4\,h\,a_y'\big]
\Bigg]
\,\,\rc\rc
&&\qquad\qquad\qquad\qquad
-E(\eta\,\cos\theta\,a'-a\,\sin\theta\,\theta')\,=\,0\,\,.
\label{eom_ax2_general_case}
\eear
It is easy to demonstrate that the equations for $A_r$, $A_{\alpha}$, and $A_{\beta}$ are trivially satisfied by our ansatz. The only non-trivial equation for the gauge field that remains to write is the one corresponding to $A_{\psi}$, which is:
\bear
&&\partial_r\Bigg[{r^2\,\sqrt{h}\,\sqrt{q^2+b^4\,a^2}\,\sqrt{(B^2+r^4)h-E^2}
\over \sqrt{\Delta}}\,a'\Bigg]-
{\sqrt{\Delta}\,\sqrt{(B^2+r^4)h-E^2}\over\sqrt{h}\, \sqrt{q^2+b^4\,a^2}}\,a\, \,\rc\rc
&&\qquad\qquad\qquad\qquad
+3r^2 a\,-\,{2(2-q)\,\eta\over b(q+\eta)}\,(B\,a_0'\,+\,E\,a_y')\cos\theta=0\,\,.
\label{eom_apsi_general_case}
\eear
Finally, one can prove that the three equations for the transverse scalars $\theta$, $\hat \theta$, and $\hat\varphi$ are the same, namely:
\bear
&&\partial_r\Bigg[{r^2\,\sin^2\theta\,\sqrt{h}\,\sqrt{q^2+b^4\,a^2}\,\sqrt{(B^2+r^4)h-E^2}
\over \sqrt{\Delta}}\,\theta'\Bigg] \rc\rc
&&\qquad\qquad
-\,{\sqrt{q^2+b^4\,a^2}\,\sqrt{(B^2+r^4)h-E^2}\over \sqrt{h}\,\sqrt{\Delta}}
\big[\Delta\,-\,b^4\,r^2\,h\,a'^{\,2}\big]
\cot{\theta} \qquad\rc\rc
&&\qquad
-(3-2b)\,q\,r^2\,\sin\theta\cos\theta\,+\,{2b^3(2-q)\over q+\eta}\,a\,\sin\theta\,
(B\,a_0'\,+\,E\,a_y')\,=\,0
\,\,.\qquad\qquad
\label{eom_theta_general_case}
\eear

Eq. (\ref{eom_ax_general_case}) allows us to eliminate $a_x'$, after which
 we have four second-order, coupled differential equations (\ref{eom_a0_general_case})-(\ref{eom_theta_general_case}) for four radial functions of $a_0$, $a_y$, $a$, and $\theta$. Solving this system in general is a quite formidable task. For this reason it is worth to look for simplifications and partial integrations. Notice that (\ref{eom_a0_general_case}) and (\ref{eom_ax2_general_case}) present some electric-magnetic symmetry. 
Actually, by combining these equations one easily finds the following constant of motion:
\beq
{r^4\,\sqrt{h}\,\sqrt{q^2+b^4\,a^2}\over \sqrt{\Delta}
\sqrt{(B^2+r^4)h-E^2}}\,\sin^2\theta\,
\big[E\,a_0'\,+h\,B\,a_y'\big]\,=\,{\rm constant}\,\,,
\label{constant_of_motion}
\eeq
which  could be used to eliminate $a_0'$ or $a_y'$ from the system of equations. Moreover, in the unflavored case ($\eta=b=q=1$), the last two terms in (\ref{eom_a0_general_case}) and (\ref{eom_ax2_general_case}) can be combined to construct the radial derivative of $a\cos\theta$, which leads to two constants of motion. In this unflavored case, $a_0$ and $a_y$  are cyclic variables and can be eliminated.

\renewcommand{\theequation}{\rm{C}.\arabic{equation}}
\setcounter{equation}{0}

\section{Kappa symmetry analysis}
\label{kappa}

The kappa symmetry matrix for a D$p$-brane in the type IIA theory is given by:
\beq
d^{p+1}\,\zeta\,\Gamma_{\kappa}\,=\,
{1\over \sqrt{- \det (g+F)}}\,e^{F}\wedge X\,\,,
\eeq
where $g$ is  the induced metric,  $\zeta^{\alpha}$ ($\alpha=0,\ldots ,p)$ are a set of worldvolume coordinates and $X$ is the polyform matrix:
\beq
X\,=\,\sum_{n}\,\gamma_{2n+1}\,\big(\Gamma_{11}\big)^{n+1}\,\,,
\eeq	
with $\gamma_{2n+1}$ being the $(2n+1)$-form whose components are the  antisymmetrized products of the induced Dirac matrices $\gamma_{\mu}$:
\beq
\gamma_{2n+1}\,=\,{1\over (2n+1)!}\,\gamma_{\mu_1\cdots \mu_{2n+1}}\,\,
d \zeta^{\mu_1}\,\wedge\cdots\wedge d \zeta^{\mu_{2n+1}}\,\,.
\eeq
In particular, we are interested in the case of a D6-brane with a flux across a (non-compact) four-cycle. The corresponding kappa symmetry matrix takes the form:
\beq
d^{7}\, \zeta\,\Gamma_{\kappa}\,=\,
{1\over \sqrt{ -\det (g+F)}}\,\Big[\,\gamma_{(7)}\,+\,F\wedge \gamma_{(5)}\, \Gamma_{11}\,+\,
{1\over 2}\,F\wedge F\wedge  \gamma_{(3)}\,+\,{1\over 6}\,
F\wedge F\wedge  F\wedge \gamma_{(1)}\,\Gamma_{11}
\Big]\,\,.
\label{Gamma_kappa_FF}
\eeq

Let us now study the conditions imposed by kappa symmetry in the case in which
the embedding is determined by the conditions (\ref{RP3-cycle}),  the worldvolume gauge field takes the form (\ref{A_full_ansatz}) with $a_x=0$, and the background is the zero-temperature supergravity solution  of \cite{Conde:2011sw}. We begin by computing  the pullbacks  of the left-invariant $SU(2)$ one-forms  $\omega^i$ of (\ref{w123}) in the $\alpha$, $\beta$, and $\psi$  variables:
\beq
\hat \omega^{1}\,=\,\hat \omega^{2}\,=\,0\,\,,\qquad\qquad
\hat \omega^{3}\,=\,2d\beta\,\,,
\eeq
whereas those of ${\cal S}^i$ and $E^i$ are:
\bear
&&\hat {\cal S}^{\alpha}\,=\,d\alpha\,\,,
\qquad
\hat {\cal S}^{1}\,=\,0\,\,,
\qquad
\hat {\cal S}^{2}\,=\,\sin\alpha\sin\theta\,d\beta\,\,,
\qquad
\hat {\cal S}^{3}\,=\,-\sin\alpha\cos\theta\,d\beta\,\,,\qquad\rc\rc
&&\hat E^{1}\,=\,\theta'\,dr\,\,,\qquad\qquad
\hat E^{2}\,=\,\sin\theta\,(d\psi+\cos\alpha\,d\beta)\,\,.
\label{pullbacks_S_E_newangles}
\eear
The pullbacks of the frame one-forms  used in Appendix B of \cite{Conde:2011sw} are:
\bear
&& \hat e^{\mu}\,=\,L\,r\,dx^{\mu}\,\,,\qquad\qquad
\hat e^{3}\,=\,{L\over r}\,dr\,\,,\qquad\qquad
\hat e^{4}\,=\,{\sqrt{q}\over b}\,L\,d\alpha\,\,,
\qquad\qquad\qquad\qquad\rc\rc
&&\hat e^{5}\,=\,0\,\,,\qquad\qquad
\hat e^{6}\,=\,L\,{\sqrt{q}\over b}\,\sin\alpha\sin\theta\,d\beta\,\,,
\qquad\qquad
\hat e^{7}\,=\,-L\,{\sqrt{q}\over b}\,\sin\alpha\cos\theta\,d\beta
\,\,,\rc\rc
&&\hat e^{8}\,=\,{L\over b}\,\theta'\,dr\,\,,\qquad
\hat e^{9}\,=\,{L\over b}\,\sin\theta\,
(d\psi+\cos\alpha\,d\beta)\,\,.
\label{pullback-es_newangles}
\eear
The corresponding  induced gamma matrices become:
\bear
&&\gamma_{x^{\mu}}\,=\,L\,r\,\Gamma_{\mu}\,\,,\qquad\qquad
\gamma_{r}\,=\,{L\over r}\,\Big(\,\Gamma_3\,+\, {r\over b}\,\theta'\,\Gamma_8\,\Big)\,\,,
\qquad\qquad
\gamma_{\alpha}\,=\,L\,{\sqrt{q}\over b}\,\,\,\Gamma_4\,\,,
\qquad\qquad
\rc\rc
&&\gamma_{\beta}\,=\,L\,{\sqrt{q}\over b}\,\sin\alpha\sin\theta\,\,
\Big[\,\Gamma_{6}-\cot\theta\,\Gamma_7+{\cot\alpha\over \sqrt{q}}\,\Gamma_9\,\Big]\,\,,
\qquad\qquad
\gamma_{\psi}\,=\,{L\,\sin\theta\over b}\,\Gamma_9\,\,.
\label{induced_gammas_newangles}
\eear

Let us next compute the different contributions on the right-hand side of (\ref{Gamma_kappa_FF}). First of all we notice that:
\beq
\gamma_{(7)}\,=\,d^7\,\zeta\,\,\gamma_*\,\,,
\eeq
where $\gamma_*$ is the antisymmetrized product of all induced gamma matrices, namely:
\beq
\gamma_*\,=\,\gamma_{t \,x\,y\, r\,\alpha\,\beta\,\psi}\,\,.
\eeq
In terms of flat 10d matrices, $\gamma_*$ can be written as:
\beq
\gamma_*\,=\,{q\over b^3}\,L^7\,r^2\,
\sin\alpha\,\sin^2\theta\,\,
\Gamma_{012}\,\,\Big(\,\Gamma_3\,+\,{r\,\theta'\over b}\,\Gamma_8\,\Big)\,
\Gamma_4\,\Big(\,\Gamma_6\,-\,\cot\theta\,\Gamma_7\,\Big)\,\Gamma_9\,\,.
\eeq
With our notation, the supersymmetric embeddings are those that satisfy $\Gamma_{\kappa}\,\epsilon\,=\,-\epsilon$, where $\epsilon$ is a Killing spinor of the background. To implement this relation we impose that $\epsilon$ satisfies the projection corresponding to a D2-brane, \ie,
\beq
\Gamma_{012}\,\epsilon\,=\,-\epsilon\,\,.
\label{D2-brane_projection}
\eeq
We also impose that $\epsilon$ satisfies the generic projections found in Appendix B of \cite{Conde:2011sw} for a generic ABJM-like geometry (Eqs.~(B.4) and (B.14) in \cite{Conde:2011sw}):
\beq
\Gamma_{47}\,\epsilon\,=\,\Gamma_{56}\,\epsilon\,=\,\Gamma_{89}\,\epsilon\,\,,
\qquad\qquad
\Gamma_{3458}\,\epsilon\,=\,-\,\epsilon\,\,.
\label{internal_projections}
\eeq
From these projections  it follows that:
\beq
\Gamma_{3469}\,\epsilon\,=\,-\Gamma_{8479}\,\epsilon\,=\,\epsilon\,\,,
\qquad\qquad
\Gamma_{3479}\,\epsilon\,=\,\Gamma_{8469}\epsilon\,=\,-\Gamma_{38}\,\epsilon\,\,.
\eeq
Using  (\ref{D2-brane_projection}) and (\ref{internal_projections}), $\gamma_*\,\epsilon$ reduces to:
\beq
\gamma_*\,\epsilon\,=\,-{q\over b^3}\,L^7\,r^2\,\sin\alpha\, \sin^2\theta \,\,
\left[\,1\,+\,{r\,\theta'\over b}\,\cot\theta\,+\,\Big(\cot\theta\,-\,{r\,\theta'\over b}\,\Big)\,
\Gamma_{38}\,\right]\epsilon\ .
\label{gamma_star_epsilon_unquenched}
\eeq

From the condition that $ \gamma_*$ acts diagonally  on $\epsilon$ (\ie, $ \gamma_* $ acts on $\epsilon$ as a matrix proportional to the unit matrix), we get the following equation for the embedding angle:
\beq
r\,\theta'\,=\,b\,\cot\theta\,\,.
\label{BPSeq_theta}
\eeq
Moreover, when (\ref{BPSeq_theta}) and the projections (\ref{D2-brane_projection}) and (\ref{internal_projections})  hold, $\gamma_{(7)}$ acts on $\epsilon$ as:
\beq
\gamma_{(7)}\,\epsilon\,=\,-d^7\,\zeta\,{q\over b^3}\,L^7\,r^2\,\sin\alpha\,\epsilon\,\,.
\eeq
Let us now study the terms in (\ref{Gamma_kappa_FF}) that are linear in the worldvolume gauge field $F$. Let us write these terms as:
\beq
F\wedge \gamma_{(5)} \Gamma_{11}\,=\,d^7\zeta\,\big[\,
\Gamma^{flux}+\Gamma^{Min}\,\big]\,\,,
\label{F_gamma5}
\eeq
where $\Gamma^{flux}$ contains the contributions of the components of $F$ along the internal directions and $\Gamma^{Min}$ is the contribution of the components of $F$ with legs along the Minkowski spacetime. It is readily verified that:
\beq
\Gamma^{flux}\,=\,
\gamma_{t\,x\,y}\,\Gamma_{11}\,\Big[\,\gamma_{\alpha\,\beta}\,F_{r\,\psi}\,-\,\gamma_{\alpha\psi}\,F_{r\,\beta}
\,+\,\gamma_{r\psi}\,F_{\alpha\,\beta}\,\Big]\,\,.
\label{Gamma_flux}
\eeq
The antisymmetric products of induced gamma matrices appearing on (\ref{Gamma_flux}) can be straightforwardly computed from (\ref{induced_gammas_newangles}):
\bear
&&\gamma_{\alpha\,\beta}\,=\,{q\over b^2}\,L^{2}\,\,\sin\alpha\,\sin\theta\,
\Big[\Gamma_{46}\,-\,\cot\theta\,\Gamma_{47}\,+\,
{\cot\alpha\over \sqrt{q}}\,\,\Gamma_{49}\,\Big]\,\,,\rc\rc
&&\gamma_{\alpha\psi}\,=\,{\sqrt{q}\over b^2}\,
L^{2}\,\sin\theta\,\Gamma_{49}\,\,,\rc\rc
&&\gamma_{r\psi}\,=\,{L^{2}\over b}\,{\sin\theta\over r}\,
\Big[\,\Gamma_{39}\,+\,{r\,\theta'\over b}\,\Gamma_{89}\,\Big]\,\,,
\label{Gamma_products_flux}
\eear
On the other hand, $\Gamma^{Min}$ is given by:
\beq
\Gamma^{Min}\,=\,L^2\,\Big(
E\,\gamma_{y r}\,-\,a_0'\,\gamma_{x y}\,+\,B\,\gamma_{t r}\,-\,a_y'\,
\gamma_{t x}\Big)\,\gamma_{\alpha\beta\psi}\,\Gamma_{11}\,\,.
\label{Gamma_Min}
\eeq
The products of the  induced Dirac matrices needed to compute $\Gamma^{Min}$ are:
\bear
&&\gamma_{y r}\,=\,L^2\,\Big[\Gamma_{23}\,+\,{r\over b}\,\theta'\,\Gamma_{28}\Big]
\,\,,\qquad\qquad
\gamma_{x y}\,=\,L^2\,r^2\,\Gamma_{12}\,\,,\rc\rc
&&\gamma_{t r}\,=\,L^2\,\Big[\Gamma_{03}\,+\,{r\over b}\,\theta'\,\Gamma_{08}\Big]
\,\,,\qquad\qquad
\gamma_{t x}\,=\,L^2\,r^2\,\Gamma_{01}\,\,,\rc\rc
&&\gamma_{\alpha\beta\gamma}\,=\,L^3\,{q\over b^3}\,\sin\alpha\sin^2\theta\,
\Big(\Gamma_{469}\,-\,\cot\theta\,\Gamma_{479}\Big)\,\,.
\label{Gamma_products_Min}
\eear
A quick inspection of the different terms appearing in $\Gamma^{flux}\epsilon$ and $\Gamma^{Min}\epsilon$  reveals that, after using the projections (\ref{D2-brane_projection}) and (\ref{internal_projections}), all terms contain products of $\Gamma$ matrices and there are no terms containing the unit matrix. Therefore, to implement the condition $\Gamma_{\kappa}\,\epsilon=-\epsilon$ we should require that
 $\Gamma^{flux}\epsilon=\Gamma^{Min}\epsilon=0$. By combining (\ref{Gamma_flux}) and (\ref{Gamma_products_flux}) we find that the product of $\Gamma$'s contained in  $\Gamma^{flux}\,\epsilon$ is:
\bear
&&{1\over L^4}\,\,\Big[\,\gamma_{\alpha\beta}\,F_{r\,\psi}\,-\,\gamma_{\alpha\psi}\,
F_{r\,\beta}
\,+\,\gamma_{r\psi}\,F_{\alpha\beta}\,\Big]\,\epsilon\,=\,
{\sin\alpha\,\sin\theta\over b}\,
\Big({q\over b}\,a'\,+\,{a\over r}\Big)
\,\Gamma_{46}\,\epsilon\,\rc\rc
&&\qquad\qquad\qquad\qquad\qquad
-{\sin\alpha\over b^2}\,
\Big(q\cos\theta\,a'\,+\,\sin\theta\,\theta'\,a\Big)\,
\Gamma_{47}\,\epsilon\,\,.
\eear
After using the equation (\ref{BPSeq_theta}) satisfied by the angle $\theta(r)$, we find that  $\Gamma^{flux}\epsilon=0$ if the flux function $a(r)$ satisfies the following first-order equation:
\beq
{a'\over a}\,=\,-{b\over q\,r}\,\,.
\label{BPSeq_a}
\eeq
When $E=B=0$ and $a_0'=a_y'=0$, Eqs.~(\ref{BPSeq_theta}) and (\ref{BPSeq_a}) guarantee that the embedding preserves two of the four supersymmetries of the background. If this is not the case, we should continue analyzing the remaining terms in $\Gamma_{\kappa}$. From (\ref{Gamma_Min}) and (\ref{Gamma_products_Min}) we get:
\bear
{1\over L^7}\,\Gamma^{Min}\,\epsilon&=&{q\over b^3}\,\sin\alpha\,\sin^2\theta\,
\left[E\left(\Gamma_{23}+{r\over b}\theta'\,\Gamma_{28} \right)+
B \left(\Gamma_{03}+{r\over b}\,\theta'\,\Gamma_{08}\right)-r^2\,a_0'\,\Gamma_{12}-
r^2\,a_y'\,\Gamma_{01}\right]\,\rc\rc
&&\qquad\qquad\qquad\qquad\qquad\qquad
\times \Big(\Gamma_{469}\,-\,\cot\theta\,\Gamma_{479}
\Big)\,\Gamma_{11}\,\epsilon\,\,.
\eear
After using the projections (\ref{internal_projections}) we can write the action of $\Gamma^{Min}$ on the Killing spinor $\epsilon$ as:
\bear
{1\over L^7}\,\Gamma^{Min}\,\epsilon &=& {q\over b^3}\,\sin\alpha\,\sin^2\theta\,
\Bigg[(E\Gamma_{2}+B\Gamma_{0})\left[1+{r\theta'\over b}\cot\theta+
\left(\cot\theta-{r\theta'\over b}\right)\,\Gamma_{38}\right]\Gamma_{11}\,\rc\rc
&&\qquad\qquad\qquad\qquad
+\Big(a_0'\,\Gamma_{2}\,-\,a_y'\,\Gamma_{0}\Big)\,
r^2\, (1+\cot\theta\,\Gamma_{38})\,\Gamma_{13}\,\Gamma_{11}\,\Bigg]\epsilon\,\,.
\eear
Using the BPS equation for $\theta$ (\ref{BPSeq_theta}), we can rewrite this last expression as:
\bear
{1\over L^7}\Gamma^{Min}\,\epsilon&=&{q\over b^3}\sin\alpha\,
\left[(E\Gamma_{2}+B\Gamma_{0})\Gamma_{11}+
\big(a_0'\Gamma_{2}-a_y'\Gamma_{0}\big)
r^2\sin^2\theta (1+\cot\theta\, \Gamma_{38})\,\Gamma_{13}\Gamma_{11}\right]\epsilon\,\,. \rc
\eear
To ensure that $\Gamma^{Min}\,\epsilon=0$ we first impose one of the following two extra projections on $\epsilon$:
\beq
\Gamma_{02}\,\epsilon\,=\,\pm\epsilon\,\,.
\label{extra_projection}
\eeq
Notice that the conditions (\ref{extra_projection}) are compatible with the projections (\ref{D2-brane_projection}) and (\ref{internal_projections}) that we have imposed so far. We get
\beq
{1\over L^7}\Gamma^{Min}\,\epsilon={q\over b^3}\sin\alpha\,
\big[(E\mp B)\Gamma_{2}\Gamma_{11}+\big(a_0'\pm a_y'\big)
r^2\sin^2\theta (1+\cot\theta\Gamma_{38})\,\Gamma_{213}\Gamma_{11}\big]\,\epsilon\,\,,
\eeq
and we have that  $\Gamma^{Min}\,\epsilon=0$ if $E$, $B$, $a_0'$, and $a_y'$   satisfy the following conditions:
\beq
E\,=\,\pm B\,\,,
\qquad\qquad
a_0'\,=\,\mp a_y'\,\,.
\label{BPS_EB}
\eeq
The two signs correspond to the two projections in (\ref{extra_projection}) (in Section \ref{BPS-sol} we have chosen the upper signs). Therefore, after imposing these conditions, we have
\beq
F\wedge \gamma_{(5)}\,\epsilon\,=\,0\,\,.
\eeq
Notice that the extra projection (\ref{extra_projection}) is only needed if the worldvolume gauge field has components along the Minkowski directions. Furthermore, one can check that the BPS equations (\ref{BPSeq_theta}), (\ref{BPSeq_a}), and (\ref{BPS_EB}) and the projections (\ref{D2-brane_projection}), (\ref{internal_projections}), and (\ref{extra_projection}) imply that the remaning terms  in $\Gamma_{\kappa}$
act on $\epsilon$ as:
\bear
&&{1\over 2}\,F\wedge F\wedge  \gamma_{(3)}\,\epsilon\,=\,-
d^7\,\zeta\,\,{b\,L^7\over q}\,r^2\,a^2\,\sin\alpha\,\epsilon\,\,,\rc\rc
&&{1\over 6}\,
F\wedge F\wedge  F\wedge \gamma_{(1)}\,\Gamma_{11}
\,\epsilon\,=\,0\,\,.
\eear
It follows that:
\beq
d^7\,\zeta\,\Gamma_{\kappa}\,\epsilon_{|_{BPS}}\,=\,-
{d^7\,\zeta\over \sqrt{-\det(g+F)}{|_{BPS}}}\,{b\,L^7\over q}\,r^2\,
\left({q^2\over b^4}\,+\,a^2\right)\,\sin\alpha\,\epsilon_{|_{BPS}}\ ,
\eeq
and one can verify by computing the DBI determinant for the BPS configuration that, indeed, $\Gamma_{\kappa}\,\epsilon_{|_{BPS}}=-\epsilon_{|_{BPS}}$. 

\vskip 1cm
\renewcommand{\theequation}{\rm{D}.\arabic{equation}}
\setcounter{equation}{0}
\medskip

\section{Fluctuations}
\label{Fluctuations}

To find the equations satisfied by the fluctuations at first order, we just compute the variation of the gauge field equations (\ref{eom_gauge_general}). One can check that the variation of $\det M$ is zero at first order and, as a consequence, the equations for the fluctuations are:
\beq
\partial_j\Big(e^{-\phi}\,\sqrt{-\det M}\,\delta{\cal J}^{ji}\Big)\,=\,\delta j^{i}\,\,.
\label{eom_fluct_gauge}
\eeq
We will restrict our attention to the case in which the only non-zero components of $\delta A$ are those along the Minkowski directions, 
\beq
\delta A\,=\,c_{\mu}(x^{\nu},r)\,dx^{\mu}\,\,.
\label{fluct_ansatz}
\eeq
Notice that in (\ref{fluct_ansatz}) we are assuming that the $c_{\mu}$'s do not depend on the internal angles. It is then easy to verify that, when the index $i$ corresponds to one of those internal directions,  the equation of motion (\ref{eom_fluct_gauge}) is satisfied automatically by the ansatz (\ref{fluct_ansatz}). Moreover, when $i=r$ this equation reduces to the following Lorentz condition:
\beq
-\partial_0\,c_0\,+\,\partial_1 c_1\,+\,\partial_2 c_2\,=\,0\,\,.
\label{Lorentz_trasn}
\eeq
Finally, when $i=\mu=0,1,2$, Eq.~(\ref{eom_fluct_gauge}) becomes:
\bear
&&{b\over q}\,\partial_r\,\Bigg(
{r^2\sin^2\theta\,\sqrt{q^2+b^4a^2}\over 
\sqrt{b^2\sin^2\theta+r^2(b^4\,a'^2\,+\,\sin^2\theta\,\theta'^{\,2})}}
\partial_r\,c^{\mu}\,\Bigg)\rc\rc
&&\qquad\qquad\qquad\qquad
+{1\over bq}\,{\sqrt{q^2+b^4a^2}\over r^2}\,
\sqrt{b^2\sin^2\theta+r^2(b^4\,a'^2\,+\,\sin^2\theta\,\theta'^{\,2})}\,
\partial^{\nu}\partial_{\nu}\,c^{\mu}\rc\rc
&&\qquad\qquad\qquad\qquad\
+{2b^2\over q}{2-q\over \eta+q}\,\big(\eta\cos\theta\,a'\,-\,a\sin\theta\theta')\,
\epsilon^{\mu\alpha\beta}\,\partial_{\alpha}\,c_{\beta}\,=\,0\,\,,
\label{eom_fluct_explicit}
\eear
where $c^{\mu}\,=\,\eta^{\mu\nu}\,c_{\nu}$ and, in our conventions, $\epsilon^{012}=1$. 
To solve these equations, let us separate variables in  $c_{\mu}(x^{\nu},r)$ as:
\beq
c_{\mu}(x^{\nu},r)\,=\,\xi_{\mu}\,e^{ik_{\nu}\,x^{\nu}}\,R(r)\,\,,
\qquad (\mu=0,1,2)\,\,,
\label{ansatz_a_fluct_c}
\eeq
where $\xi_{\mu}$ is a constant polarization vector. It follows immediately  from (\ref{Lorentz_trasn}) that this vector satisfies the transversality condition (\ref{transversality}).

In order to write the fluctuation equation for the radial function $R$ in a compact form, let us define the differential  operator ${\cal O}$, which acts on any function of the radial coordinate $R(r)$ as:
\bear
&&{\cal O}\,R\,\equiv\,{b\over q}\,
\partial_r\left[\,
{r^2\sin^2\theta\,\sqrt{q^2+b^4a^2}\over 
\sqrt{b^2\sin^2\theta+r^2(b^4\,a'^2\,+\,\sin^2\theta\,\theta'^{\,2})}}
\partial_r\,R
\right] \rc\rc
&&
\qquad\qquad\qquad\qquad\qquad
+{m^2\over bq}\,{\sqrt{q^2+b^4a^2}\over r^2}\,
\sqrt{b^2\sin^2\theta+r^2(b^4\,a'^2\,+\,\sin^2\theta\,\theta'^{\,2})}\,\,R \ ,
\qquad\qquad
\eear
where $m$ is the mass of the dual meson (see (\ref{mass_mesons})). 
We also define  the function $\Lambda(r)$ as:
\beq
\Lambda(r)\,\equiv\,
{2b^2\over q}{2-q\over \eta+q}\,\big(\eta\cos\theta\,a'\,-\,a\sin\theta\theta')\,\,.
\eeq
Then, the fluctuation equation can be written as:
\beq
\xi^{\mu}\,{\cal O}\,R\,+\,i\epsilon^{\mu\alpha\beta}\,
k_{\alpha}\,\xi_{\beta}\,\Lambda R\,=\,0\,\,.
\label{Minkowski_fluct_eqs}
\eeq
Moreover, by substituting the values of the functions $\theta(r)$ and $a(r)$ which correspond to a SUSY embedding (\ref{theta_SUSY}) and (\ref{a_SUSY}), we can greatly simplify the operator ${\cal O}$ and the function $\Lambda$. We get:
\bear
&&{\cal O}\,R\,=\,\partial_r\left[r^{2-2b}\big(r^{2b}-r_*^{2b}\big)\partial_r\,R\right]\,+\,
{m^2\over r^{2(3-b)}}\,\big(r^{2(2-b)}+(2-b)^2\,b^2\,Q^2\,r_*^{2(2-b)}\big)\,R\,\,,\rc\rc
&&\Lambda\,=\,2b\,(2-b)\,(4-3b)\,Q\,{r_*^2\over r^3}\,\,.
\label{calO_Lambda_explicit}
\eear
The three equations in  (\ref{Minkowski_fluct_eqs}) are coupled to each other. Let us see how they can be decoupled and reduced to a single ordinary differential equation. First of all, without loss of generality we pick the Minkowski momentum as $k^{\mu}\,=\,(\omega, k, 0)$ with  the  meson mass being  $m=\sqrt{\omega^2-k^2}$. The transverse polarization has been written in (\ref{polarization}) in terms of two unknown constants $\xi_1$ and $\xi_2$. For this parametrization  of $\xi_{\mu}$ one can
show that the equations for $\mu=0$ and $\mu=1$ in   (\ref{Minkowski_fluct_eqs})  are equivalent and that the remaining two equations are just:
\bear
&& \xi_1\,{\cal O}\,R\,+\,i\omega \xi_2\,\Lambda R\,=\,0\,\,,\rc\rc
&& \xi_2\,{\cal O}\,R\,-\,i\omega\left[1\,-\,{k^2\over \omega^2}\right]\,\xi_1\,
\Lambda R\,=\,0\,\,.
\eear
To decouple these equations, let us consider the complex combinations $\chi_{\pm}$ defined in (\ref{chi_pm}). Then, one can  straightforwardly show that the system (\ref{Minkowski_fluct_eqs}) can be reduced to the equations:
\beq
\chi_{+}\,\Big({\cal O}\,R\,+\,m\,\Lambda\,R\Big)\,=\,0\,\,,
\qquad\qquad
\chi_{-}\,\Big({\cal O}\,R\,-\,m\,\Lambda\,R\Big)\,=\,0\,\,.
\label{chi-equation}
\eeq
Obviously, $\chi_{\pm}$ can be eliminated from (\ref{chi-equation})  when they are non-vanishing  and the system can be reduced to two ordinary differential equations for the radial functions $R_{\pm}$, which can be written as:
\beq
\partial_r\,\big[r^{2-2b}(r^{2b}-r_*^{2b})\partial_r R_{\pm}\big]+
\Big[{m^2\over r^{2(3-b)}}\,\Big(r^{2(2-b)}+(2-b)^2\,b^2\,Q^2\,r_*^{2(2-b)}\Big)\,\pm\,
m\,\Lambda(r)\Big]R_{\pm}=0\,\,.
\label{R-eq}
\eeq
To find the mass spectrum we must compute the values of $m$ leading to a normalizable solution. This can be done numerically by the shooting technique.  We present these numerical results for the two types of modes  in Figs.~\ref{mass_spectrum} and \ref{mass_spectrum_unquenched}. 

\begin{figure}[ht]
\center
 \includegraphics[width=0.47\textwidth]{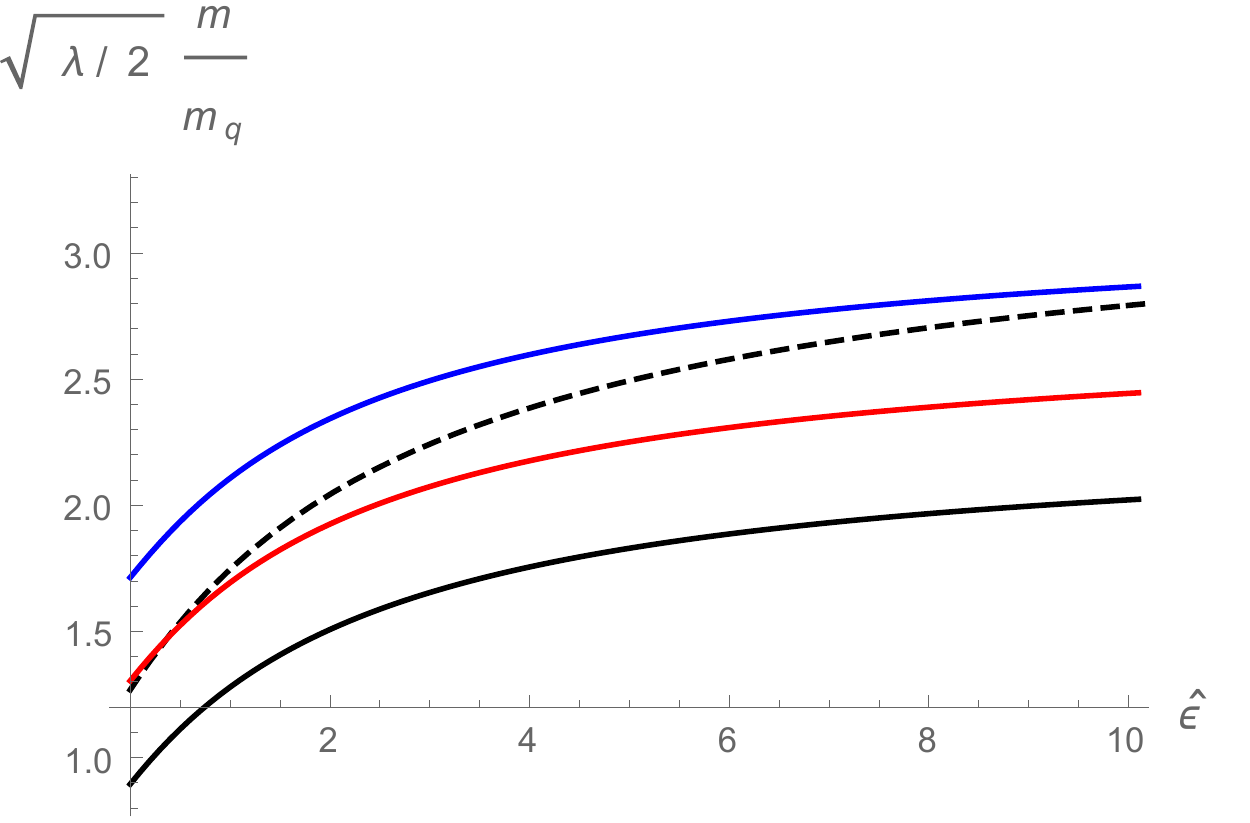}
 \qquad
 \includegraphics[width=0.47\textwidth]{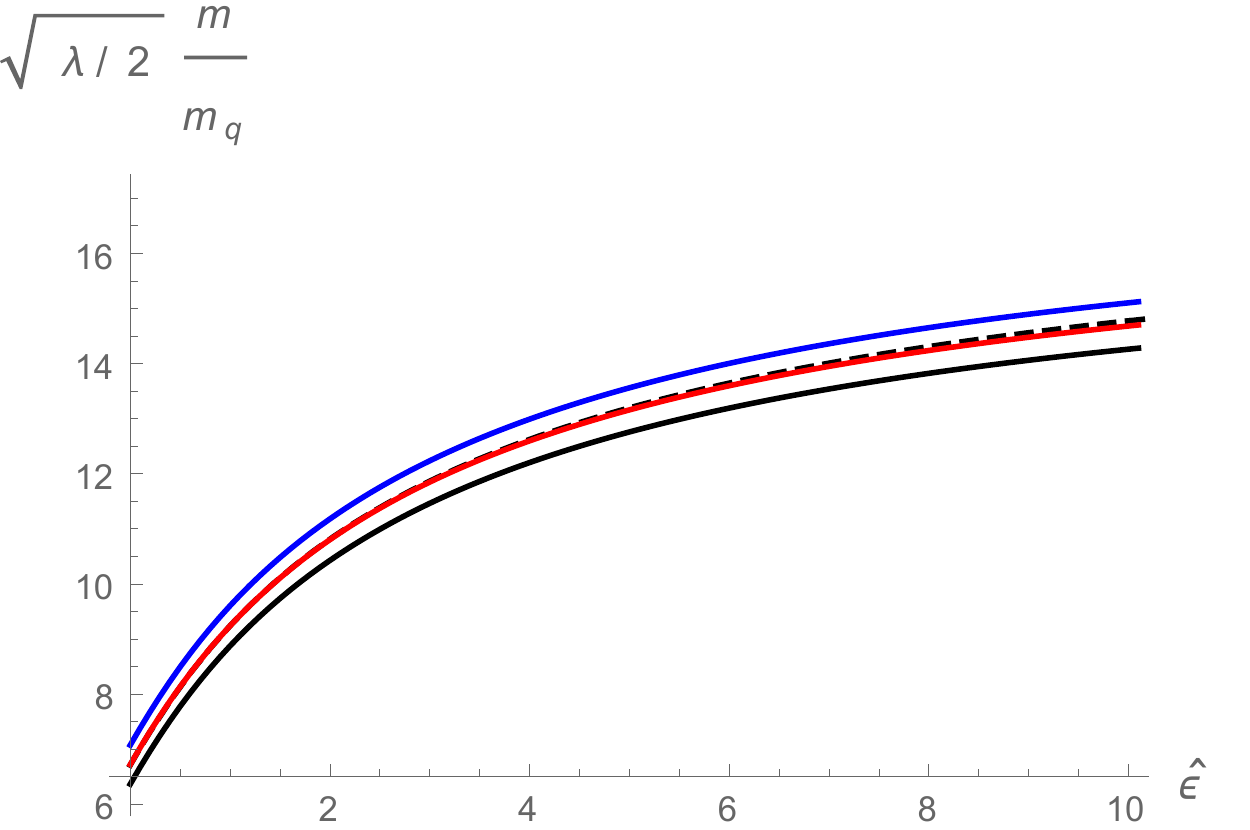}
 \caption{Meson masses in the unquenched background as a function of the flavor deformation parameter $\hat \epsilon$ for $\sqrt\lambda M/N=1$ and two values of the excitation integer: $n=0$ (left) and $n=3$ (right).  The upper blue (lower black) curve corresponds to the mode $\chi_-$ ($\chi_+$). The intermediate red curve is the average of the two curves and the dashed black curve is the WKB estimate (\ref{WKB_masses_general}).}
 \label{mass_spectrum_unquenched}
\end{figure}

\subsection{WKB mass spectrum}

When the mass $m$ is large we can neglect the term containing the function $\Lambda$ in the fluctuation equation (\ref{R-eq}), and we can estimate the mass levels by using the WKB method developed in \cite{RS}. Indeed,  let us consider a differential equation of the form
\beq
\partial_{r}\,\big(\,f(r)\,\partial_{r}\,R\,\big)\,+\,
 m^2\,h(r)\,R\,=\,0\,\,,
\label{ODE}
\eeq
where $ m$ is the mass parameter and $f(r)$ and $h(r)$  are two
arbitrary functions that are independent of $ m$. We will assume that
near $r\approx r_*$ and $r\approx \infty$ these functions behave as:
\bear
&&f\approx f_1(r-r_*)^{s_1}\,\,,
\,\,\,\,\,\,\,\,\,\,\,\,\,\,
h\approx h_1(r-r_*)^{s_2}\,\,,
\,\,\,\,\,\,\,\,\,\,\,\,\,\,{\rm as}\,\,r\to r_*\,\,,\rc\rc
&&f\approx f_2\,r^{r_1}\,\,,
\,\,\,\,\,\,\,\,\,\,\,\,\,\,
h\approx h_2\, r^{r_2}\,\,,
\,\,\,\,\,\,\,\,\,\,\,\,\,\,{\rm as}\,\,r\to \infty\,\,,
\label{coeff_exp_wkb_def}
\eear
where $f_i$, $h_i$, $s_i$, and $r_i$ are constants. Then, the mass levels for large quantum number $n$ can be approximately written in terms of these constants as \cite{RS}:
\beq
 m^2_{WKB}\,=\,{\pi^2\over 
\xi^2}\,(n+1)\,\bigg(n\,+\,{|s_1-1|\over s_2-s_1+2}+{|r_1-1|\over r_1-r_2-2}
\bigg)\,\,,\quad\quad \quad (n\ge 0)\,\,,
\label{generallWKB}
\eeq
where $\xi$ is the following integral:
\beq
\xi\,=\,\int_{r_*}^{\infty} dr \,\,\sqrt{{h(r)\over f(r)}}\,\,.
\label{xi-WKBintegral}
\eeq
In our case  $f$ and $h$ are the functions:
\bear
&&f(r)\,=\,r^{2-2b}(r^{2b}-r_*^{2b})\,\,,\rc\rc
&&h(r)\,=\,{1\over r^{2(3-b)}}\,\Big(r^{2(2-b)}+
(2-b)^2\,b^2\,Q^2\,r_*^{2(2-b)}\Big)\,\,.
\eear
The behavior of these functions at $r=r_*$ is characterized by the following values of the coefficients and exponents defined in (\ref{coeff_exp_wkb_def}):
\bear
&&f_1\,=\,2\,b\,r_*\,\,,\qquad\qquad\qquad\qquad\qquad
 s_1\,=\,1\,\,,\rc\rc
&&h_1\,=\,{1+(2-b)^2\,b^2\,Q^2\over r_*^2}\,\,,\qquad\qquad\,\,
 s_2\,=\,0\,\,.
\eear
Similarly, for the behavior at large $r$ we obtain:
\bear
&&f_2\,=\,1\,\,,\qquad\qquad \qquad\qquad \qquad\qquad\qquad
r_1\,=\,2\,\,,\rc\rc
&&h_2\,=\,1\,\,,\qquad\qquad \qquad\qquad \qquad\qquad\qquad
  r_2\,=-2\,\,.
\eear
Therefore, the WKB mass spectrum is:
\beq
m_{WKB}\,=\,{\pi\over \sqrt{2}\,\xi(b,Q)}\,\sqrt{(n+1)(2n+1)}\,\,,
\label{WKB_masses_general}
\eeq
where $ \xi(b,Q)$ is the following integral:
\beq
\xi(b,Q)\,\equiv\,{1\over r_*}\,
\int_1^{\infty}\,{dz\over z^{2(2-b)}}\,
{\sqrt{z^{2(2-b)}\,+\,(2-b)^2\,b^2\,Q^2}\over
\sqrt{z^{2b}-1}}\,\,.
\eeq
By expanding in series the square root in the numerator and integrating term by term, we can express $ \xi(b,Q)$ as the following series:
\beq
\xi(b,Q)\,=\,-{1\over 4\,b\,r_*}\,
\sum_{p=0}^{\infty}\,(-1)^p\,
\big[(2-b)\,b\,Q\big]^{2p}\,
{\Gamma\Big(p-{1\over 2}\Big)\over  p!}\,\,
{\Gamma\Big({1+2p(2-b)\over 2b}\Big)\over 
\Gamma\Big({1+2p(2-b)\over 2b}+{1\over 2}\Big)}\,\,.
\eeq
Some particular values of the integral  $ \xi(b,Q)$ are:
\bear
&& \xi(b=1,Q)\,=\,{\pi\over 2 r_*}\,\,F\Big(-{1\over 2}, {1\over 2}; 1;-Q^2\,\Big)
\,\,,\rc\rc
&& \xi(b,Q=0)\,=\,{\sqrt{\pi}\over r_*}\,\,
{\Gamma\Big({2b+1\over 2b}\Big)\over 
\Gamma\Big({b+1\over 2b}\Big)}\,\,.
\eear
Interestingly, for $b=1$ and $Q=0$ (the unflavored model without internal flux) the WKB formula for the mass levels is exact. Moreover, for large $Q$ we can approximate $\xi(b,Q)$ as:
\beq
\xi(b,Q)\,\approx\,{(2-b)\,b\,Q\over r_*}\,
\int_1^{\infty}\,{dz\over z^{2(2-b)}
\sqrt{z^{2b}-1}
}\,=\,{\sqrt{\pi}\,Q\over r_*}\,{(2-b)\,b\over 3-b}\,
{\Gamma\Big({3+b\over 2b}\Big)\over 
\Gamma\Big({3\over 2b}\Big)}
\,\,.
\eeq
It follows that, for fixed quantum number $n$,  the WKB mass levels  for large $Q$ decrease as $1/Q$ according to the equation:
\beq
m_{WKB}\,\approx\,{\sqrt{\pi}\,r_*\over \sqrt{2}\,Q}\,{3-b\over (2-b)\,b}\,
{\Gamma\Big({3\over 2b}\Big)\over \Gamma\Big({3+b\over 2b}\Big)}\,
\sqrt{(n+1)(2n+1)}\,\,.
\eeq
In Fig.~\ref{mass_spectrum_unquenched} we compare the WKB estimates using (\ref{WKB_masses_general})  and the numerical results. The WKB method, however, is not valid at large values of $Q$, as it falls off the validity regime of \cite{RS}. Our numerical studies verified this expectation.

\end{document}